\documentclass{iopjournal}

\usepackage{amssymb,amsmath}
\usepackage{mathtext}
\usepackage{float}
\usepackage{graphicx,colordvi}
 \usepackage{colordvi}
\usepackage{color}
\usepackage{calc}
\usepackage{ifthen}
\usepackage{epsfig}
\usepackage[toc,page]{appendix}
\usepackage{booktabs}
\usepackage{makecell}
\usepackage{anyfontsize}
\usepackage{orcidlink} 
\usepackage{ulem}
\def\be{\begin{equation}}
\def\ee{\end{equation}}
\def\bea{\begin{eqnarray}}
\def\eea{\end{eqnarray}}
\def\l{\label}

\def\hahat{\hat{H}}

\def\hahat0{\hat{H}_0}

\def\cos{\hbox{cos}}
\def\sin{\hbox{sin}}

\def\MItf{\mathcal{T}}

\def\d{\hbox{d}}
\def\eps{\varepsilon}
\def\epsi{\mathcal{E}}

\def\siml{\hbox{\kern.1em \lower.6ex \hbox{$\sim$} \kern-1.12em
 \raise.6ex \hbox{$<$} \kern.1em}}
\def\simg{\hbox{\kern.1em \lower.6ex \hbox{$\sim$} \kern-1.12em
 \raise.6ex \hbox{$>$} \kern.1em}}

\begin{document}

\articletype{Paper} 

\title{ROTATING NEUTRON STARS 
  WITHIN THE MACROSCOPIC EFFECTIVE-SURFACE
  APPROXIMATION}

\author{A.G.~Magner$^{1,*}$\orcidlink{0000-0002-7702-7282}, 
S.P.~Maydanyuk$^{2,3}$\orcidlink{0000-0001-7798-1271},
A.~Bonasera$^{4}$\orcidlink{0000-0002-7702-7282},
H.~Zheng$^{5}$\orcidlink{0000-0001-5509-4970}, 
S.N.~Fedotkin$^{1}$\orcidlink{0000-0002-5953-6278},
A.I.~Levon$^6$\orcidlink{0000-0002-9880-1927},
T.~Depastas$^4$\orcidlink{0000-0002-6458-4418}, 
U.V.~Grygoriev$^1$\orcidlink{0000-0002-2684-2586},
and A.A.~Uleiev$^1$\orcidlink{0009-0009-7543-4608}} 

\affil{$^1$Nuclear Theory Department, Institute for Nuclear Research, Kyiv 03028, Ukraine}

\affil{$^2$Nuclear Processes Department, Institute for Nuclear Research, Kyiv 03028, Ukraine}

\affil{$^3$Southern Center for Nuclear-Science Theory (SCNT), Institute of
      Modern Physics, Chinese Academy of Sciences, Huizhou 516000, China}

\affil{$^4$Cyclotron Institute, Texas A\&M University,
      College Station, Texas 77843, USA}

\affil{$^5$School of Physics and Information Technology,  Shaanxi Normal
      University, Xi'an 710119,  China}

\affil{$^6$Heavy Ions Physics Department, Institute for Nuclear Research, Kyiv 03028, Ukraine}

\email{alexander.magner66@gmail.com}

\keywords{nuclear astrophysics, neutron stars, compact stars,
  extended Thomas Fermi, effective  surface, adiabatic rotations}

\begin{abstract}
 The macroscopic model for a neutron star
 (NS) as 
 a finite perfect fluid at
 the equilibrium 
 is extended to rotating systems
 by incorporating the linear perturbation expansion 
 over a small frequency $\omega$
 near Schwarzschild outer-inner gravitational metric within the  
 effective-surface (ES)
 approach. 
 The NS angular momentum $I$
 and moment of inertia (MI) 
 for a slow stationary
 azimuthal rotation around 
 the symmetry axis are calculated by using 
 the Kerr metric approach in
 spherical coordinates, and compared with
 Boyer-Lindquist (outer) and Hogan
 (inner) metric results.
 The volume and gradient-surface terms of the
 macroscopic NS energy density 
 $\mathcal{E}(\rho)$ (Equation of State)
 are taken into account 
 at the leading order of  the leptodermic parameter $a/R \ll 1$,
 where $a$ is the ES crust thickness and
 $R$ is the NS effective radius.
 The analytical macroscopic NS MI expressions,
 $\Theta=\d I/\d \omega=\tilde{\Theta}/(1-\mathcal{T}_{t\varphi})$, 
 have been obtained in terms of the
 statistically averaged MI, $\tilde{\Theta}$,
 and its time and 
 azimuthal-angle
 $t,\varphi$  correlation, 
 $\mathcal{T}_{t\varphi}$, as
 sums of the 
 volume and surface 
 components.
 The 
 MI $\Theta$ is changed significantly  
 as function of the effective radius $R$ because
 of  a strong gravity. We found the 
 additional constraint for the NS radius to 
 smaller 
 accessible ranges which is due mainly to  
 the
 $t,\varphi$  correlations and surface contributions.
 The adiabaticity conditions 
 for applicability of the linear
 perturbation theory is carried out 
 for 
 several neutron stars 
 with a strong gravity and relatively large rotation periods.
 \end{abstract}

\section{Introduction}
\label{introd}

Recently, the simultaneous data 
on masses and radii 
of neutron stars (NSs)
were
obtained
\cite{JN17,BA19,BA19-1,GR19,TR19,MM19,MM19-1,CC20,GR19-1,TR21,MM21,VD22,RK23,AD24,YK24,DS24,DC24,MM25,LM25,wiki:PSR-Cen-X-3,YK26}
with a good accuracy 
for the double rotating neutron stars;  see also  early successful data presented in Refs.~
\cite{TG10,SN11,MR11,TG13,MF15,MB13,DK15,TMT17},
    in particular, for the rotational periods.
This is obviously a second
new big progress to improve and test the theoretical description 
    of NSs
based on
the independent data on masses and radii beyond the models for the Equation
of States (EoSs).

As is well known,
 Tolman suggested 
 \cite{RT39} first to study a neutron star 
 considering it as a perfect 
 dense fluid of finite sizes at its
equilibrium under
the gravitational, nuclear and other realistic fields;
see also his book \cite{RT87}, chapt. 7, sect. 96,  based on the
    Schwarzschild solution \cite{KS16out,KS16in} 
        for the sphere of incompressible fluid.
Within this spherical model, the equations of the
General Relativity Theory (GRT)
have
     been 
 solved by Schwarzschild  in terms of
    the outer and inner parameters,
    $\lambda$ and $\nu$, of the 
   time-space element
    (see
also Ref.~\cite{LLv2}, chapts. 11 and 12 for the exterior branch)
\be\l{Schwarz}
   \!\! {\rm d} s^2\!=\!e^\nu c^2{\rm d} 
   t^2\!-\!e^\lambda{\rm d} r^2\!-\!
     r^2 {\rm d}\theta^2\!-\!
     r^2 \sin^2\theta {\rm d} \varphi^2,
     \ee
    For the formulation of a complete system
    of equations for the pressure $\mathcal{P}(r)$
    as function of the radial coordinate $r$ Tolman, suggested
    to derive also
    the EoS energy density, $\epsi=\epsi(\rho)$,  
       where $\rho$ is
    the density of stellar matter. 
The EoS should be found 
for a perfect finite-dense fluid under realistic interactions.
Following Tolman's ideas, Oppenheimer and Volkoff (TOV)
 have derived \cite{OV39} the simple
equations by using essentially
the macroscopic stellar properties; see 
 Ref.~\cite{LLv6} (chapt.~1, sects.~1-3; chapt.~7, sect.~61;
chapt.~15, sect.~133) and 
those of Ref.~\cite{RW82} (chapt.~1)
     for such
    systems. 
As shown in 
Ref.~\cite{RT87}, one can derive even the analytical solution
for the pressure $\mathcal{P}(r)$ 
by using the step-like density and Schwarzschild inner gravity
    solution \cite{KS16in} for a uniform 
    incompressible-fluid sphere.
    The numerical solutions to the TOV equations
    were trusted also for continuously sharp
behavior
of the density.
So far, these 
equations \cite{OV39} were solved with  
the EoS, $\epsi=\epsi(\rho)$, which was sometimes different 
from the assumptions
used in the TOV derivations. In particular, 
 a popular polytropic EoS 
\cite{LLv5} (chapt.~11),
\cite{ST04} (chapt.~9), and \cite{HPY07} (chapt.~6)
    is similar to that for
    a particle gas system\footnote{The gas system
        assumes
    a long-range inter-particle interaction,
     in contrast to the dense
    fluid (or amorphous solid)
    system where a mean distance between particles is of the order 
    of the interaction range \cite{LLv6}.
    }
    with, however, fitted parameters 
    under the baryon, lepton and gravitational fields
    \cite{LLv2}.
  
             Another important issue used in the TOV derivations
        is the boundary conditions for
        a continuous transformation of the
        outer to inner Schwarzschild metric near a 
        finite  NS radius $R$.
        The strong
        gravity leads to the restrictions
        which are imposed on the NS radius $R$.
 By using the TOV equations derived for
a dense
    fluid system 
        with the boundary conditions mentioned above,
    the NS mass $M$ can be obtained 
    as function of the
NS radius $R$, $M=M(R)$, with well-known
important Schwarzshild constraints. 
 Following these ideas, one can
 assume for enough large neutron stars within the
     simplest statistical macroscopic
 approximation \cite{MM24,MM25npa} that 
 the mean NS density  $\rho(r)$ is almost
  constant $\overline{\rho}$ inside NS
  with a continuous sharp
  change within a small but finite diffused  crust region having 
  a thickness $a$, $a/R\ll 1$, where $R$
  is a mean NS radius.   Another restriction used below in
  our analytical derivations 
   for the inner densities larger than the nuclear matter
  density $\rho^{}_0 = 10^{14}$ g/cm$^3$, is
  $1 \siml \overline{\rho}/\rho^{}_0\siml 4$. This restriction
  leads approximately to 
  the upper limit  for NS masses $M$,
  being larger than or of the order of the Sun mass
  $M_\odot$, namely  $M\siml 3 M_{\odot}$; 
      see details in the text.
  
    We note the advantage
    of a simple macroscopic treatment taking into account a
    finiteness of the realistic
    NSs.
  We should point out also
    the importance of the gradient terms in the local
    energy density $\epsi(\rho)$
    for a macroscopic condition of the system equilibrium
     in terms
    of the local characteristics 
    (the pressure $\mathcal{P}$ and the density $\rho$)
    near the NS surface,
       in spite of a
       relatively small NS crust thickness $a \ll R$
        \cite{MM25npa}.
    These statistically averaged NS peculiarities 
        are hinted fruitfully by the
        recent observational 
        data for masses and radii, mentioned above,
        and many theoretical
        works,   
      e.g.,  Refs.~\cite{HPY07,BBP71,BPS71,WF88,CB97,AA98,PH00Poland,LP01,HP01PRL,HP01PRC,SH06,Ko08,CH08,ZB11,FCPG13,PFCPG13,SG13,RW14,BZ14,BC15,GB18,LH19,KM21prc,KM21aj,GN23,SBL23,Pe23,FG23,DFG23,LJ23,SS23,Pe24,CM25,XV24}.
    Clear specific definitions and complete updated results
    for the energy density with
    density gradient (surface) terms and 
     bulk dense  matter (inner) properties
  with many inter-particle
 interactions in the non-relativistic and relativistic cases
can be found in the recent reviews; see
Refs.~\cite{GB18,SBL23}.

 As the TOV equations
    \cite{OV39} used along with
    the EoS in a lot
    of astrophysical works
     on the NS,
    it would be logical to
    agree the arguments for the
    specific derivations of both
    these equations. This is the main motivation of our and TOV
    macroscopic approaches which is important also 
    for studying the NS rotations.
         Within nuclear astrophysics 
    we have to take into account
   the strong gravity
    as in 
        the derivations
    of the TOV equations,
    in contrast to the nuclear physics. Our suggestion for 
        these coordinations
        \cite{MM25npa} does not exclude more microscopic
    EoSs. However, it requires 
    the corresponding essential
    modification of the TOV 
    equations; see Refs.~ 
        \cite{MM24,MM25npa}.

Taking Tolman's ideas, we 
first extended \cite{MM24,MM25npa} the
 EoS
 to those for such a finite dense-macroscopic \cite{LLv2,LLv6}
 system of particles, in more details for the case of
 non-rotating NS systems. 
 In this leptodermic system, one finds
 the
 density $\rho$ as function of the
 radial coordinate $r$ with
exponentially decreasing behavior 
from an almost constant saturation
value inside the dense system
to that through the NS effective
surface (ES) within relatively a small inner-crust range $a$.
The ES is defined 
as the points of spatial coordinates with a maximum of density gradients.
To obtain the analytical solutions for the
particle density and EoS, we 
use the
 effective surface (leptodermic) 
 approximation (ESA),  $a/R \ll 1$, for sufficiently heavy NSs; see
Refs.~\cite{ST04,HPY07,BPS71,PH00Poland,LP01,HP01PRL,HP01PRC,FCPG13,PFCPG13,GB18,LH19,KM21prc,KM21aj,SBL23,Pe23,FG23,DFG23,Pe24,XV24}.
In our macroscopic approach, the NS radius $R$ is
the curvature radius 
of the NS
ES. 
Within the
ESA, the
simple and
accurate solutions of many 
realistic problems involving the 
density $\rho$ distributions 
were obtained  
   for heavy nuclei
   \cite{strtyap,strmagbr,strmagden,MS09,BM13,BM15},
   dense molecular systems \cite{RW82,vdW},
       metallic clusters (electronic Fermi-liquids) \cite{FG01,BB03},
and neutron stars; see, e.g., Refs.~\cite{MM24,MM25npa}.
The ESA exploits
the property of saturation of the statistically averaged
density $\rho$
inside of a 
known
dense system such as
molecular, nuclear  or mesoscopic Fermi-liquids
systems.
A saturation is a characteristic macroscopic 
feature 
\footnote{
    For the 
    dense molecular (e.g., liquid-drop) systems,
    van der Waals (vdW) \cite{vdW} suggested the phenomenological
capillary theory
which predicted the results for the  
density
$\rho$ and surface tension coefficients $\sigma$;
see also Ref.~\cite{RW82}.
}
 of liquid drops (amorphous solids), nuclei,
 and presumably, NSs.
The realistic energy-density distribution 
is minimal at a certain saturation
density 
of particles (nucleons or neutrons) 
in the infinite 
  nuclear  matter. As a result, 
  relatively narrow edge region exists in finite nuclei or 
      rotating neutron star crust
in which the macroscopic
density of the stellar matter drops exponentially
from its almost central value to zero. We assume
here that the statistically averaged density part 
inside of 
a large dense system far from the ES can be relatively changed
a little. 
 This saturation property was found in many
    macroscopic finite-dense systems; see, e.g.,
     \cite{RT87,LLv6,RW82,HPY07,BPS71,BB03}.
      The equilibrium condition means that 
      the variation of
      the total energy $E$ over the
density $\rho$ is zero under the constraints which fix 
some
integrals of
motion beyond the energy $E$ by the Lagrange method. The Lagrange multipliers
  are determined by these constraints
  within the local energy-density theory, in particular,
  the extended Thomas-Fermi 
  (ETF) approach, 
  well known 
  from nuclear and mesoscopic metallic-cluster physics; see 
      Refs.~\cite{brguehak} and \cite{BB03} (chapt.~4).
   The Lagrange
  equilibrium equations 
  can be reduced 
  to a
  simple one-dimensional 
  catastrophe equation for the 
  density $\rho$ in the leading 
  normal-to-ES direction;
   see Appendix~\ref{appA} for details. 
Such an equation mainly determines approximately the density
distribution 
across the diffused surface layer of the relatively small order,
$a/R \ll 1$, 
in the body-ﬁxed coordinate system, i.e. 
with zero rotational 
frequency $\omega$.  
A small leptodermic parameter, $a/R$, related (self)consistently with
    the inter-particle interaction and incompressibility 
        \cite{MM24,MM25npa,AM25npae},
in the 
expansion within the ESA  can be used
for analytically solving the variation problem
for a minimum of the system energy 
with constraints for
a fixed particle number, and other integrals of motion,
 such as angular momentum,
 quadrupole deformation, etc. The remarkable Schwarzschild two-branch
     (outer-inner) solution, Refs.~\cite{KS16out,KS16in,RT87},
     for the GRT metric is the zero order approach within the
     ESA (a step-function density). The first order ESA corrections
     stand for
     the NS surface contributions. 
When this edge distribution of the 
density is known, the leading
static and dynamic density distributions which 
correspond to the
diffused surface conditions can be easily constructed. 
 Relatively a large change of the density $\rho$ on a small
distance $a$ with respect to 
the ES curvature radius $R$ (i.e., the existence of
    the effective surface itself)
takes place for 
any kind of fluid-matter drops.
Inside of such dense systems,
the density $\rho$ is changed 
slightly around
a mean 
inner-density constant $\overline{\rho}$ 
relatively far from the ES. Therefore,
one obtains
essential effects
of the surface capillary pressure of the general 
    statistical van der Waals
theory \cite{RW82,vdW}. 
    This surface pressure contribution plays a significant
    role, along with the volume inner pressure component, which are
    both much influenced
by a strong gravity, for the macroscopic equilibrium condition of rotating
NSs.

The accuracy of the ESA was checked in 
Ref.~\cite{strmagden} for the 
nuclear physics problems
by comparing the results 
with the existing nuclear theories like Hartree-Fock (HF)
\cite{vauthbrink} and ETF \cite{BB03,brguehak}, based on the 
Skyrme forces 
\cite{CB97,brguehak,vauthbrink,skyrme,barjac,ringshuk,blaizot,krivin,CB98,gramvoros79,gramvoros80}, 
but for the simplest
case without spin-orbit and asymmetry terms 
of the energy density functional.
The extension of the ESA
to the nuclear isotopic symmetry 
and spin-orbit interaction has been done in
Refs. \cite{MS09,BM13,BM15}; see also Refs.~\cite{KS01,KS08,KS13}.
The Swiatecki derivative
terms of the symmetry energy for
heavy nuclei
\cite{danielewicz2,vinas09,vinas09-1,vinas10,vinas10-1,vinas11,vinas14,Pi09}
were
taken into account  within the ESA in Ref. \cite{BM15}. The
discussions of the progress in nuclear physics and astrophysics
within the relativistic local density approach,
can be found, e.g., in reviews
Refs.~\cite{GB18,SBL23,NVR11}; see also intensive studies in
Refs.~\cite{Pe23,CH08,Pi21}. 
         The significant ES corrections to the TOV equations
        \cite{RT87,OV39} for neutron stars
        have been derived analytically in Ref.~\cite{MM25npa}.

 We are ready now to formulate
 the macroscopic NS rotational problem, as a small 
 rotational-energy perturbation
 of the strongly gravitating spherically-symmetric 
 background described 
 by the outer and inner Schwarzschild metric. The basic ideas for  
            studying a slow rotating spherical
           system in the GRT
           was largely formulated already in
            Ref.~\cite{LT18} by Lense and
            Thirring by 
           using an extended 
            Schwarzschild metric.
            The frequency $\omega$ dependence of 
           the gravitational metric for 
         a slow rotating 
        star was
        first more consequently
         obtained within the GRT 
         by  
        Kerr
        in Ref.~\cite{RK63}. 
              Simple clear derivations of the Kerr metric approach
        were discussed 
        \cite{LLv2,Teu15,SSMV24,MV08} for a relatively 
        slow rotation
         by accounting also for small (quadrupole or spheroidal)
        deformations
        of the gravitating system.                
        The simplest form of the Kerr metric
        in terms of 
        other more 
      transparent variables
           was found for the region outside of the 
        NS ($r>R$) by Boyer and Lindquist
        \cite{BL67}, and 
        for the inner NS part ($r<R$) 
        by Hogan \cite{PH76,PC78}; see
        also Refs.~\cite{LLv2,Teu15}. Independently, the  
        specific perturbation  formulation based on the non-rotating
        Schwarzschild gravitational metric and taking into  account
        small multipole-surface deformations
        due to a  small uniform
        frequency-fluid  rotation
        was suggested by Hartle and Thorne in Refs.~\cite{JH67,HT68};
        see also Ref.~\cite{AK78}. Their approach  
        for the description of the
    uniformly rotating neutron stars
    with a constant angular velocity $\omega$ 
    is developed in Refs.~\cite{MV08,FIP86,NS03,AW08,NS13,NS17}
    for a small angular momentum of
    the perfect stellar fluid. 
      
In the present work, we extend the ES approximation
of Refs.~\cite{MM24,MM25npa,strmagden,MS09,BM13,BM15}, 
to the 
rotating neutron stars for their slow macroscopic
rotational
motion within the linear perturbation theory (LPT).
 The Kerr metric background 
  for rotating neutron stars
    by using the Boyer and
    Lindquist variables \cite{BL67} and 
    Hogan gravitational metric \cite{PH76,PC78}
    at small rotation frequencies
    $\omega$ is shown in Sect.~\ref{KMAM}.
        Then, we derived the 
        solutions of the 
        GRT equations
    in the linear perturbation approach over the rotational frequency
    $\omega$, and compared with the Boyer-Linquist and Hogan
    approaches in Sect.~\ref{GRTsol}.
    The macroscopic EoS for
    the gravitational energy density and NS masses are
    considered 
    in Sect.~\ref{MEOS}.
 In Sect.~\ref{SFA}, the adiabatic
NS moment of inertia (MI) for small rotational frequencies
with the surface and $t,\varphi$ correlation corrections
 were derived. 
         The results of our
        calculations
        and comparison with 
       observational  data are discussed in
        Sect.~\ref{discres}. The main results
    are summarized in Sect.~\ref{concl}.
        Useful details of calculations
are shown in
 Appendixes \ref{appA}-\ref{appD}.

 \section{The Kerr metric and  NS angular momentum approach}
 \l{KMAM}
 \subsection{The Kerr metric}
 \l{KBL}

      Space-time metric solutions to the
        GRT
        equations for the uniformly rotating
        NS around the symmetry axis
         at 
        a small rotational frequency $\omega$
        by taking into account a frequency and
        deformation dependence of the 
        gravitational
        field was suggested by Kerr \cite{RK63,Teu15}, 
            see also Hartle and
        Thorne approach \cite{JH67,HT68}.  
        By using a transformation to a more 
        clear spherical 4-coordinate system,
        $\{t,r,\theta,\varphi\}$,
        the Kerr metric solution 
        \cite{RK63} was specified 
        by
        Boyer and Lindquist
        \cite{BL67} for outer region, $r>R$
             (see also Refs.~\cite{LLv2,Teu15}),
  \be\l{BLm}
 \d s^2 =\left(1 -\frac{r^{}_gr}{\Sigma}\right)
  c^2\d t^2 +\frac{2r^{}_g~r \Omega}{\Sigma}\sin^2\theta~c\d t\d \varphi
  -\frac{\Sigma}{\Delta}\d r^2
-\Sigma \d \theta^2 
- \left(r^2+\Omega^2+
\frac{r_gr\Omega^2}{\Sigma}\sin^2\theta\right)
 \sin^2 \theta \d \varphi^2.
 \ee
 where $r^{}_g$ is the following gravitational radius:
        \be\l{rg}
        r^{}_g=\frac{2MG}{c^2}~,
        \ee
  Here, $M$ is the NS mass,
           $G$ is the gravitational constant, $c$ is the speed of light,
 \be\l{SD}
 \Sigma=r^2+\Omega^2\cos^2\theta,\quad \Delta=r^2-r_gr+\Omega^2~,
  \ee
  $\Omega$ is the
  rotational 
  frequency parameter
   (``$a$'' in the notations of
 Refs.\ \cite{RK63,BL67,LLv2}),
 $\Omega \propto \omega$, 
 and $\omega$ is  the 
  azimuthal rotational frequency.
  Asymptotically far from the gravitating masses, where we may
  neglect density $\rho(r)$,
  one has
 the relation 
  between the parameter $\Omega$ 
  and the NS angular momentum
 $I$,
 \be\l{OmegaK}
 I \approx \omega\Theta 
 \approx \Omega M c~,
 \ee
 where $\Theta$ is the NS MI. It is convenient to introduce
the dimensionless rotation-frequency parameter
\cite{LB11,PP12},
 \be\l{rotpar}
 \overline{\omega}=\frac{\Omega c^2}{M G}
 \approx \frac{Ic}{M^2G}\approx
 ~\frac{\Theta c}{M^2G}~\omega~.
 \ee
 Our perturbation approach means the linear approximation
     over the dimensionless
     frequency (or angular momentum) parameter $\overline{\omega}$.
 The solution, Eq.~(\ref{BLm}), 
 has the outer ($r>R$)
 Schwarzschild metric limit (\ref{Schwarz}) for
 $\Omega \rightarrow 0$ \cite{RT87,LLv2}:
     \be\l{smlim}
   \d s^2
  \rightarrow \left(1 - \frac{r^{}_g}{r}\right)
 c^2\d t^2 -\left(1-\frac{r^{}_g}{r}\right)^{-1}
 \d r^2
  - r^2 \left(\d \theta^2 +
  \sin^2 \theta \d \varphi^2\right).
  \ee
  For the outer Schwarzschild metric space-time line element,
  Eq.~(\ref{Schwarz}),
one has $\nu$ and $\lambda$ at $r>R$ 
in the leading leptodermic
approximation, 
(see Eq.~(\ref{smlim}) and Ref.~\cite{RT87}), 
\be\l{nulamout}
\nu_{\rm out}=\ln\left(1-\frac{r^{}_g}{r}\right),\quad
\lambda_{\rm out}=-\nu_{\rm out}~. 
\ee

For the inner Kerr metric $g^{}_{\mu\nu}$,
 one can use 
the expression 
suggested by Hogan 
\cite{PH76,PC78} 
in the following
form:
\be\l{inkerrmet}
\d s^2=f c^2\d t^2 +2(1-f)\Omega\sin^2\theta c\d t\d \varphi
-\left[(1-f)\Omega^2\sin^4\theta+(r^2+\Omega^2)\sin^2\theta\right]\d
\varphi^2
-\frac{\Sigma}{\chi}\d r^2 -\Sigma\d \theta^2,
\ee
where
\bea\l{fchiQ}
\chi&=&r^2-qQ^2\Sigma+\Omega^2,\qquad
f=\left(\frac32\sqrt{1-qR^2}-\frac12\sqrt{1-qQ^2}\right)^2,\nonumber\\
Q&=&\frac{\Sigma}{r},\qquad q=\frac{r^{}_g}{R^3},\quad\mbox{for}\quad R>r^{}_g~.
\eea
Here, $r^{}_g$ and $\Sigma$ are given by Eqs.~(\ref{rg}) and (\ref{SD}),
respectively. 
The metric, Eq.~(\ref{inkerrmet}), for $\Omega \rightarrow 0$ 
turns into the inner 
Schwarzschild metric 
\cite{RT87} at $r < R$,
     \be\l{SchwarzIN}
        {\rm d} s^2=c^2\left[A -
       B\sqrt{1- r^2/R^2_{\rm S}}\right]^2 {\rm d} t^2
      -\frac{{\rm d} r^2}{1-r^2/R^2_{\rm S}}-
     r^2 {\rm d}\theta^2 -
     r^2 \sin^2\theta {\rm d} \varphi^2~, 
     \ee
     where
 \be\l{Tolconsschwm}
      A=\frac32\sqrt{1-\frac{R^2}{R_{\rm S}^2}},\qquad
      B=\frac12~,
      \ee
          and $R_{\rm S}$ is another Schwarzschild 
          radius\footnote{The
          notations for the Schwarzschild 
          radius $R_{\rm S}$, used in the book \cite{RT87}, is $R$, 
          but the effective radius is denoted there as
 $r^{}_1$. To avoid a confusion,
  one should mention also that
 sometimes the terminology ``Schwarzschild radius" is used in the 
 literature for
 $r^{}_g$, 
 Eq.~(\ref{rg}), that is not the case in this paper.},
        \be\l{RS}
        R_{\rm S}=\sqrt{\frac{3 c^4}{8\pi G
            \overline{\epsi}}},\quad \overline{\epsi}=\epsi(\overline{\rho})
         \approx \overline{\rho}c^2~,
        \ee
       and $ \overline{\rho}$  is the inner saturation density 
         at 
         an equilibrium  \cite{RT87}. 
             As seen,
         the gravitation radius $r_g$, Eq.~(\ref{rg}), will be the lower,
         and $R_{\rm S}$, Eq.~(\ref{RS}), will stand for  the
         upper boundary for the
         NS radius. Here,
         the leptodermic property of the almost constant
    density $\rho(r)$ inside,
        and relatively a sharp continuous decrease in a small diffused
    crust region with a finite thickness 
    $a$ outside of the NS, $a/R\ll 1$,
    has been assumed.
 The  condition for
 matching of the inner and outer
 Schwarzschild metrics (Refs.~\cite{KS16in,RT87},
 was really used in the zero
     leptodermic approach,
\be\l{boundcond1}
\frac{R}{R_{\rm S}}=\sqrt{\frac{r_g}{R}}~.
\ee
For 
the inner solutions $\nu$ and $\lambda$, one finds from
    Eq.~(\ref{SchwarzIN})
at zero approximation over the rotation perturbation:
\be\l{nulamin}
\nu_{\rm in}=2\ln\left(A-B\sqrt{1-\frac{r^2}{R_{\rm S}^2}}\right),\qquad
\lambda_{\rm in}=-\ln\left(1-\frac{r^2}{R_{\rm S}^2}\right).
\ee

 \subsection{The NS angular momentum}
 \l{AM}

 For the angular momentum $I$ in the case of the stationary rotation
of a perfect fluid system around its symmetry axis
under a strong gravitational field,
one writes \cite{FIP86,NS03,AW08,NS13,NS17}
\be\l{Igen}
I=\frac{1}{c^2}\int T_\mu^{~\nu}~\phi^\mu \hat{n}_\nu  
\d \mathcal{V}_{\Omega}~.
\ee
Here, $ T_{\mu}^{~\nu}$ is the energy-momentum tensor of the
perfect dense fluid in the  statistical
generating
free-energy ensemble \cite{LLv6,LLv5,FIP86,NS13},
\be\l{Tmn}
T_{\mu}^{~\nu}=\mathcal{E}(\mathcal{\rho}) u_\mu u^\nu +\mathcal{P} g_{\mu}^{~\nu},
\quad u_\mu=u^{t}\left(t_\mu +\omega \phi_\mu\right).
\ee
where $\mathcal{E}(\rho)$ is the energy density, and  $\mathcal{P}$ is
the pressure.
In Eqs.~(\ref{Igen}) and (\ref{Tmn}), the Killing vectors,
$t_\mu\propto \omega$, and
$\phi^\mu$ are defined 
as that acting in
the azimuthal direction reflecting axial
symmetry, and $\hat{n}_\nu$ is the normal vector to the
spatial volume-hyper surface in the 4-space.
     The
     particle velocity $u_\mu$ 
     is normalized to one, $u_\mu u^\mu=1$, and
     $u^t=e^{-\nu/2}/(1-v^2/c^2)^{1/2}\approx e^{-\nu/2}$ 
         for small rotational-particle
     velocities $v$~.
     Using simple algebraic transformations and
     Eqs.~(\ref{Igen}), (\ref{Tmn}), (\ref{OmegaK}),
     and the definitions for the
     Killing vectors 
     (see Refs.\cite{NS03,NS13}) in the
     first order approximation over $\overline{\omega}$, for the moment
     of inertia $\Theta$ one obtains
\be\l{Theta2}
\Theta=\frac{\partial I}{\partial \omega}=
\frac{1}{c^2}\int \epsi(\rho)e^{-\nu} r_\perp^2\left(1+
\frac{2\tau}{r^2}\frac{\Theta}{M}\right)\d \mathcal{V}~.
\ee
The first component in
brackets is related to the global
     rotation as a whole, as for the rigid  body. The second part
     is proportional to the off-diagonal correlation element of the
     gravity metric $g^{}_{t \varphi} \propto \omega\tau\Theta$.
     The solution
     of the GRT equation for $\tau(r)$ 
         is derived in Appendix \ref{appB}, and
     discussed below.
Notice that the pressure component of the energy-momentum tensor
$T_{\mu}^{~\nu}$, Eq.~(\ref{Tmn}),  does not contribute into Eq.~(\ref{Theta2})
for $\Theta$.
Up to quadratic terms, the integration in Eq.~(\ref{Theta2})
    is carried out over
the spatial system volume $\d \mathcal{V}$,
\be\l{dV}
\d\mathcal{V}^{}_\Omega\approx \d\mathcal{V}=
\mathcal{J}(r) r^2 \sin \theta \d r \d \theta \d \varphi,
 \ee
 where $\mathcal{J}(r)$ takes into account the gravity,
\bea\l{Jin}
\mathcal{J}(r)&=& \frac{1}{\sqrt{1-r^2/R_{\rm S}^2}}~~~(r\le R)\\
&=&\frac{1}{\sqrt{1-r^{}_g/r}}~~~(r > R)~,
\l{Jout}
\eea
Solving Eq.~(\ref{Theta2}) with respect to $\Theta$, with
the definition Eq.~(\ref{rg}) for the gravitational
radius $r^{}_g$,
one obtains
\be\l{MI}
\Theta=\frac{\partial I}{\partial \omega}=
\frac{\tilde{\Theta}}{1 - \mathcal{T}_{t\varphi}}~,
\ee
Here, $\tilde{\Theta}$ is the statistically 
averaged MI, 
\be\l{MIt}
\tilde{\Theta}=\frac{1}{c^2}
\int  \epsi(\rho)e^{-\nu} r_\perp^2\d \mathcal{V}~.
\ee
and $\mathcal{T}_{t\varphi}$ is the  
time and azimuthal-angle, $t,\varphi$, 
correction
 due to
the off-diagonal gravitational 
metric element $g_{t\varphi}\propto \tau(r)$,
\be\l{MItf}
\mathcal{T}_{t\varphi}\approx \frac{2}{Mc^2}
\int \epsi(\rho)e^{-\nu} \frac{\tau}{r^2}
r_\perp^2 \d \mathcal{V}~.
    \ee
    Thus, owing to this gravitation self-consistency,
    $I \Rightarrow \omega \Rightarrow I$,
        Eqs.~(\ref{Theta2}) and
    (\ref{OmegaK}),
one arrives at the pole MI expression (\ref{MI}).
Note that this pole structure
can be proved within the perturbation 
    theory which includes also the
$\omega^2$
corrections; see
Refs.~\cite{BL67,PH76,JH67,HT68} but the expressions for
the counterparts $\tilde{\Theta}$ and $\MItf_{t\varphi}$ are expected to be
more complicate. The pole in Eq.~(\ref{MI}) 
  is
  the root
  $R_{\rm K}$ of the function $\mathcal{T}_{t\varphi}(R)-1$, if it exists,
  \be\l{RKpole}
\mathcal{T}_{t\varphi}(R_{\rm K})-1=0~.
  \ee
  This pole is the reason of an 
     additional constraint on the NS radius $R$ because of the
     rotational perturbations.
 In Eqs.~(\ref{MIt}) and (\ref{MItf}),
$\nu$ is defined by the gravitational metric,
Eqs.~(\ref{BLm}) and (\ref{SchwarzIN}), in the form 
(\ref{AWmet}) for the linear perturbation approach
over 
rotational 
frequency $\overline{\omega}$ [Eq.~(\ref{rotpar})],
and
$r^{}_\perp=r~\sin\theta$.

    Using the statistical averaging over the canonical ensemble
\cite{LLv5}, one
can exclude the off-diagonal correlation term,
$\propto \d t\d \varphi$, from the gravitational metric,
Eq.~(\ref{AWmet}). Removing therefore 
the second term, $\propto \tau$, from
parentheses of Eq.~(\ref{Theta2}), for the MI one arrives at
Eq.~(\ref{MIt}) for the averaged MI $\tilde{\Theta}$. 
 This approach is physically close to  
that used in Ref.~\cite{AK78} where
 the off-diagonal correlation term,
$\propto \d t\d \varphi$, has been neglected.  
However, for small rotation frequencies $\omega$, the 
 gravitational field is approximated by that of the 
three-dimensional 
spheroidal deformations.
 Notice also
 that the density $\rho$
in Eq.~(\ref{MIt}) for the MI
within the leading leptodermic
approximation over the parameter $a/R$ is 
independent of $\Omega$
at linear order, $\Omega \propto \omega \propto \overline{\omega}$
(Appendix \ref{appA}).

\section{The GRT solutions for Kerr metric}
\label{GRTsol}
\subsection{Outer $r>R$ off-diagonal metric element}
\l{OUT}

As shown in Appendix \ref{appB} within the linear approximation over
$\overline{\omega}$, Eq.~(\ref{OmegaK}), on 
the 
Schwarzschild metric background, 
for the outer region ($r>R$)
one arrives at Eq.~(\ref{14taux.1}).
Then, one obtains
  \be\l{14tauxsolgen}
  \tau=\alpha^{}_1 x^2+\frac{\alpha^{}_2}{x} ~,\qquad x=\frac{r}{r^{}_g}~,
  \ee
  where $\alpha^{}_1$ and $\alpha^{}_2$ are arbitrary integration constants.
  Using the boundary condition on infinity
  ($\alpha^{}_1=0$), one obtains 
 \be\l{14tauxsol.1}
 \tau= \frac{\alpha^{}_2r^{}_g}{r}~.
    \ee
    For $\alpha^{}_2=1$,
   this solution
   is identical to that obtained  in Refs.~\cite{BL67,MV08}; see
   Eq.~(\ref{tauout}).

     \begin{figure}
       \begin{center}
         \includegraphics[clip,width=8.0cm]{Fig1_soleqtau.eps}
         \end{center}

        \vspace{-0.6cm}
    \caption{{\small
        Numerical inner solutions of the off-diagonal
        GRT
        equation (\ref{14tauin}) for the key quantity
        $\tau(r)$ of Kerr metric in the linear perturbation approach
        are shown as functions of the radial variable $r$ in units of the
        Schwarzschild
        radius $R_{\rm S}$, Eq.~(\ref{RS}), $x=r/R_{\rm S}$,
        $r\leq R$,
        by different lines
        (logarithmic scales for both axes).
        Arrows show the end values of
               $R/R_{\rm S}=0.1$ and $0.5$.
        For the same arbitrary
        initial conditions, $\tau(0)=0$ and
       $\tau^\prime(0)=0$, 
       all $\tau(x)$ are calculated
       in terms of the Mathematica
       ``InterpolatingFunction'' solution
       of Eq.~(\ref{14tauin}) for $\tau(x)$.
         }} 
  \label{fig1}
       \end{figure}

     \subsection{Inner $r\leq R$ solutions}
     \l{IN}

     Substituting Eq.~(\ref{nulamin}) into
   Eq.~(\ref{14tau}), one obtains Eq.~(\ref{14tauin}) for $\tau(x)$,
   where $x=r/R_{\rm S}$ 
       [see Eq.~(\ref{RS}) for $R_{\rm S}$]. Figure \ref{fig1}
   shows the numerical 
   solutions
   of this equation with the initial conditions $\tau(0)=0$ and
   $\d\tau(0)/\d x=0$ for the inner range of the radial variable, $r \leq R$.
   As seen from this figure, one finds almost the
   linear dependence of $\ln \tau$
   on  $\ln x$.
   For the natural condition
   $\tau(0)=0$ we obtain a parallel shift of the lines if variating the
   initial
   derivative in these plots. The end point $r=R$ for each curve
   depends of the specific properties of the NS, - its mass $M$ and radius
   $R$, - through the boundary condition (\ref{boundcond1}) in terms of
   the ratio $R/R_{\rm S}$.
     As far as we consider
   here  $R/R_{\rm S}$ as a variable, these universal
   plots are independent of
   the 
    observational data for NS  masses and radii.
 Notable discrepancies are found for larger values of $R/R_{\rm S} \simg 0.9$.
Taking into account that $\tau(x)$
should be integrated over $x$ till the NS
surface $r=R$, for instance
in calculations of the
moments of inertia $\Theta$, Eq.~(\ref{MI}),  such a discrepancy
is not very important for the 
   final results.

Expanding then
   the coefficients $a_k(x)$ of Eq.~(\ref{14tauin})
   over $R/R_{\rm S}$ at $R/R_{\rm S}\ll 1$, and therefore,
    over $x=r/R_{\rm S}$ at $r \leq R$ up to second order, one arrives at
   Eq.~(\ref{14tauinexp}) for $\tau(x)$. This equation has the solutions,
   Eq.~(\ref{14tauinexpsolgen}),
   in terms of the Bessel functions. Using the condition of finiteness
   of $\tau(x)$ inside the NS (for $x\rightarrow 0$),
   one arrives at 
\be\l{14tauinexpsol}
\tau(x)=c^{}_1 \sqrt{x}  J_{p}\left(\sqrt{5}x\right)~,\qquad x=r/R_{\rm S}~,
\ee
where
\be\l{p}
p=\sqrt{9-12\xi^2}/2~,\qquad \xi=R/R_{\rm S},
\ee
and $c^{}_1$ is another arbitrary integration constant.
The Bessel function solution, Eq.~(\ref{14tauinexpsol}),
    is related
    with a good accuracy to the linear part up to about
    $R/R_{\rm S}=0.5$ in Fig.~\ref{fig1}.    
    As mentioned above, the discrepancies between this approximate
    but analytically given Bessel-function solution,
Eq.~(\ref{14tauinexpsol}), and exact numerical values
for $\tau(x)$ 
at $R/R_{\rm S} \simg 0.9$ in the calculations of the MI
are expected to do not influence much on the final results; see more
discussions below.
For small $R/R_{\rm S} \siml 0.1$
       and 
           all presented lines, one finds
       the 
       Bessel solution (\ref{14tauinexpsol}) for $\tau$
      to be close to that of the
        approximate equation
        (\ref{14tauinexp}).
                Notice that using the analytical
result, Eq.~(\ref{14tauinexpsol}),
one should remember the upper restriction on  the NS mass
$M$ approximately by $3 M_\odot$, as mentioned in Introduction.

The power expansion of the solution, Eq.~(\ref{14tauinexpsol}),
up to fourth order terms has the form:
\be\l{14tauinexpsolexp4}
\tau(x)=
\frac{5^{p/2} x^{(1+2p)/2}}{4\left(1+p\right)
  2^{p}\Gamma(1+p)}
\left[4(1+p)-5x^2+O\left(x^4\right)\right]c^{}_1~.
\ee
Expanding this equation 
over $\xi$ through $p$, Eq.~(\ref{p}), one finds 
\be\l{14tauinexpsolexp4x2R2}
\tau(x)= \alpha^{}_0c^{}_1x^2 + ...,
\qquad
\alpha^{}_0=\frac{10-\psi(5/2)}{3~\sqrt{2\pi\sqrt{5}}}\approx 0.826765...~,
\ee
where
$\psi(z)$ is the digamma function (or the polygamma function
$\psi^{(n)}(z)=\d^{n+1}\ln\Gamma(z)/\d z^{n+1}$ at $n=0$,
$\psi(5/2)=0.703157...$),
and dots in Eq.~(\ref{14tauinexpsolexp4})
show the terms of higher
(fourth) order
(as $\sim \xi^2x^2 \ln(x),
\xi^2x^2, x^4$, and so on).

The integration constant $c^{}_1$, Eq.~(\ref{14tauinexpsol}),
can be found approximately from the boundary condition,
Eq.~(\ref{boundcond1}),
for slitching the inner solution,  Eq.~(\ref{14tauinexpsol}), 
with the outer result,
Eq.~(\ref{14tauxsol.1}), at 
$\alpha^{}_2=1$, 
in accordance with the results of
Refs.~\cite{RT87,BL67},
on the ES at $r=R$ in the leptodermic approximation, $a/R \ll 1 $.
The continuous outer-inner transformation of the off-diagonal
element of the gravitational metric is also carried out 
    to obtain 
\be\l{c1}
c^{}_1
=
\frac{r_g}{R\sqrt{R/R_{\rm S}}  J_{p}\left(\sqrt{5}R/R_{\rm S}\right)}~,
\ee
where
$r^{}_g$ and $R_{\rm S}$ are the gravitational 
[Eq.~(\ref{rg})] and the other [Eq.~(\ref{RS})]
    Schwarzschild radii,
and $p$ is given by Eq.~(\ref{p}). For small $R/R_{\rm S}$, one has
from Eqs.~(\ref{c1})  and (\ref{14tauinexpsolexp4x2R2})
\be\l{c1exp}
c^{}_1=\frac{r^{}_gR_{\rm S}^2}{R^3\alpha^{}_0}
\left[1+O\left(\frac{R^2}{R_{\rm S}^2}\right)\right]~,
\ee
where $\alpha^{}_0$ is the constant given by Eq.~(\ref{14tauinexpsolexp4x2R2}).
Using the boundary condition for
    slitching the outer to inner Schwarzschild solutions at zero order
    approximation, Eq.~(\ref{boundcond1}), one simply finds
    \be\l{c1appr0}
  c^{}_1\approx \frac{1}{\alpha^{}_0}
  \left[1+O\left(\frac{R^2}{R_{\rm S}^2}\right)\right]
  \approx 0.83\! 
  \left[1+O\left(\frac{R^2}{R_{\rm S}^2}\right)\right]~.
\ee
Thus, for the inner coefficient $\tau$ with Eq.~(\ref{c1exp}) for $c^{}_1$,
one obtains approximately
\be\l{tauapprox}
\tau(r) \approx \frac{r^2}{R_{\rm S}^2}
\left[1+O\left(\frac{r^2}{R_{\rm S}^2}\right)\right]
= \frac{2MG r^2}{R^3c^2}
\left[1+O\left(\frac{r^2}{R_{\rm S}^2}\right)\right]~,
\qquad r\leq R ~.
\ee
This quadratic approximation obviously obeys the
    boundary condition on the
NS surface, $r=R$, Eq.~(\ref{boundcond1}). For $a \ll R$ we may consider this
linear approximation
because the corrections to this approach for the gravitational metric are of
the second order which are neglected in our calculations.

Notice also that for Hogan's approach, Eqs.~(\ref{inkerrmet})
and (\ref{nulamin}), for $\tau$
one has 
\bea\l{tauin}
\tau&=&
1-\left(A-B\sqrt{1-\frac{r^2}{R_{\rm S}^2}}\right)^2~,
\qquad r\le R~,\\
  &=&\frac{r^{}_g}{r},\qquad r>R~.
  \l{tauout}
\eea
Expanding Eq.~(\ref{tauin}) at $r<R$ over the parameter, $\xi=R/R_{\rm S}$,
and then, over the variable, $x=r/R_{\rm S}$, up to second order,
one writes
 \be\l{tauinexpHog}
 \tau=\frac32 \xi^2-\frac{r^2}{2R_{\rm S}^2} +
 O\left(\frac{r^4}{R_{\rm S}^4}\right)~,\quad r<R~.
 \ee
    This asymptotic expression
    coincides with our
   approximate result, Eq.~(\ref{tauapprox}), at the ES, $r=R$.
    Our more accurate result for the correlation coefficient
   $\tau(r)$, Eq.~(\ref{14tauinexpsol}), differs  
    essentially from  Hogan's corresponding expression, Eq.~(\ref{tauin}).
    For instance, Eq.~(\ref{tauinexpHog})
   has a finite limit
   in the NS center, in contrast to the
   zero  limit in our derivations; see
   critical comments for the Hogan metric in Ref.~\cite{PC78}.
   However, as shown below, the moments of inertia $\Theta$ which
   are determined by the integrals, Eqs.~(\ref{MI})-(\ref{MItf}),
   and calculated in Ref.~\cite{AM25npae},
   are qualitatively  close to those obtained
   below (Sect.~\ref{NSMI}) for the cases of importance of MIs
   surface contributions, 
       and sufficiently far away from their poles,
   in spite of a significant quantitative difference 
       near these poles (see more discussions below).

\section{Macroscopic equation of state}
\label{MEOS}

For calculations of the MI, Eqs.~(\ref{MI})-(\ref{MItf}), 
we need the macroscopic energy density $\epsi(\rho)$
at zero frequency $\omega$. 
The total macroscopic energy $E$ 
can be written in terms of $\epsi(\rho)$ 
as \cite{strmagden,BM15}
\be\l{energytot}
E=\int \epsi \left(\rho({\bf r})\right)\;\d \mathcal{V},
\ee
where
\be\l{enerden}
\epsi\left(\rho\right) =\mathcal{A}(\rho)
+\mathcal{B}(\rho)\left(\nabla \rho\right)^2~,
\ee
 and $\mathcal{A}(\rho)$
and $\mathcal{B}(\rho)$ are smooth 
functions of the density $\rho$. 
 They are coefficients in expansion 
of the energy density over gradients of
$\rho$ in the ETF approach.
A non-gradient part $\mathcal{A}(\rho)$ of the energy density
$\epsi$ 
can be presented\footnote{In Ref.~\cite{MM24},
  the particle number density $n=\rho/m$ with the effective
  test-particle mass $m$ is used instead of the mass density
  $\rho$ of the present version. Then,
   one has
  $\epsi_0=\epsi(\overline{n})
  \approx -b_V^{(G)}\overline{n}$, where $b_V^{(G)}$ is the separation energy per
  particle,  $\overline{n}=\overline{\rho}/m$.
  } (at $\omega=0$) as
\be\l{vol0}
\mathcal{A}=\rho c^2+\eps(\rho)
+ \mathcal{U}(\rho)~, 
\ee
where
\be\l{eps}
\eps(\rho)=\frac{K}{18 m \overline{\rho}^2}~
\rho\left(\rho-\overline{\rho}\right)^2~,
\ee
and  $m$ is the effective test-particle mass.
The third
term $\mathcal{U}$ in Eq.~(\ref{vol0}) takes into account
 the statistically (macroscopically)
    averaged gravitational contribution
\cite{MM24}.

As well known \cite{RT87,LLv2}, the 
GRT equations
in the non-relativistic
limit and/or for a weak
 gravitational ﬁeld can be transformed to the 
Poisson continuity equation for the gravitational potential $\Phi$.
As shown in Ref.~\cite{RT87}, in this limit,
$g^{}_{00}=e^\nu$ of a more general Schwarzschild metric,
Eq.~(\ref{Schwarz}), can be presented for small
$\nu$ as $e^\nu \approx 1+\nu =1 +2\Phi/c^2$,
where $\Phi$ is the solution of the Poisson equation \cite{LLv2},
 $\nu\approx 2\Phi/c^2$. We
will use a more general deﬁnition, based on the gravitational
component $\mathcal{U}(\rho)$ of the macroscopic energy 
 density
[see Eqs.~(\ref{enerden}) and (\ref{vol0})]. 
Due to a large speed of light $c$, 
this component asymptotically approaches
$\mathcal{U}= 
-\rho c^2\nu/2$,
    where $\nu$ is determined by the
    Schwarzschild metric solution of the GRT, Ref.~\cite{RT87}.
    Then, 
    on the one hand,
    in the Newtonian limit, the energy density 
    per unit of density $\rho$ 
        (with the opposite sign),
    $-\mathcal{U}/\rho$,
    will be asymptotically equal to the gravitational potential
$\Phi$.
 On the other hand, we may  
 use 
 our macroscopic approach 
 for $\mathcal{U}(\rho)$
within the leptodermic
approach. Therefore, after such a 
statistical averaging,
the gravitational part
$\mathcal{U}$ of the macroscopic energy density
$\epsi(\rho)$ can be
considered as
function of the radial coordinate
 $r$ through the
density
$\rho=\rho(r)$. Then,
 for leptodermic  systems, statistically averaged
    over NS particle number,
    accounting for the gravitational
component of the incompressibility, one can expand 
    $\mathcal{U}(\rho)$
over powers of
the difference $\rho-\overline{\rho}$
   near
the mean  
density
$\overline{\rho}$, inside of the ES, up to 
second order,
\be\l{phiexp}
\mathcal{U}(\rho)=
\mathcal{U}_0+\frac12~\mathcal{U}^{}_2~(\rho-\overline{\rho})^2~,
\ee
where 
$\mathcal{U}_0=\mathcal{U}(\overline{\rho})$
and
$\mathcal{U}_2=
\partial^2 \mathcal{U}(\overline{\rho})/\partial \rho^2$ 
is the second derivative of the $\mathcal{U}(\rho)$ over $\rho$  
    taken at the 
    internal constant density $\overline{\rho}$.
   A linear term
disappears because of 
the condition of
minimum of the energy per particle for
an isolated system;  see Ref.~\cite{MM24}.
 With the 
expansion, Eq.~(\ref{phiexp}),
for $\mathcal{A}(\rho)$,
Eq.~(\ref{vol0}), one obtains   
\be\l{vol}
\mathcal{A}=\rho c^2 +\eps^{}_G(\rho)~,
\ee
where
  \be\l{epsKG}
  \eps^{}_G(\rho)=\eps(\rho)+ 
\frac{\mathcal{U}_2}{2m} \rho(\rho-\overline{\rho})^2
=\frac{K_G}{18m \overline{\rho}^2}~
\rho\left(\rho-\overline{\rho}\right)^2~,
\ee
and $\eps(\rho)$ is given by 
    Eq.~(\ref{eps}).
The total incompressibility modulus is modified by the
gravitational field ($K_G>0$),
\be\l{KG}
K_G=K+9 \overline{\rho}^2\mathcal{U}_2~,
\ee
where $K$ is the non-gravitational (e.g., nuclear) 
incompressibility
    component.
    Notice that our second order approximation for
    expansion of the gravitational energy-density
    component  
    $\mathcal{U}$, Eq.~(\ref{phiexp}), 
    agrees consistently with
    the energy density presentation 
        for any large dense-finite system because
    of their leptodermic property
    up to second order
    gradients, Eq.~(\ref{enerden}).
    The latter is, in turn, rather general, in
    particular,
    for all known Skyrme forces in nuclear physics taking
    into account the volume and surface terms.
   Another famous example is the 
        van der Waals forces for a large  dense finite
        molecular system 
       as a liquid drop
    \cite{RW82}.

  The coefficient $\mathcal{B}(\rho)$ in front of
the gradient squared term
in Eq.~(\ref{enerden}) is approximated 
    for simplicity  by a constant \cite{MM24},  
\be\l{surf}
\mathcal{B}(\rho) = \mathcal{C}+\mathcal{D}\rho +
\frac{\Gamma}{\rho} \approx \mathcal{C}~,
\ee
    These terms are associated with the nuclear Skyrme interaction modified by
the gravitation component. First term is related to the
interaction term which is a main reason of
the diﬀuse surface thickness
for a non-rotating NS. In this sense it has a more general meaning
including the
main eﬀective interaction in a dense molecular system, studied by van der
Waals \cite{RW82,vdW}; see also the footnote 2 in page 3.
Therefore, we will call this component as that of
the vdW-Skyrme interaction. The second term is coming from the spin-orbit
interaction. This interaction might be important, for instance, for
any dense rotating
fluid systems with a sharp surface edge, i.e., with large
density-gradient terms
in the surface region;
see also the
same general arguments for using the ETF approach in
Refs.~\cite{BB03,brguehak}.
The last term is a gradient
correction to the kinetic energy ($\hbar^2$ correction of the 
kinetic energy in the general
ETF approach \cite{BB03,brguehak}); see the Wilets result for the 
density $\rho$ in 
Ref.~\cite{MM24}.
Thus, the forms Eq.~(\ref{vol}) 
for $\mathcal{A}$ and Eq.~(\ref{surf}) 
for $\mathcal{B}$ are rather general among the
simplest analytical solutions for any dense ﬁnite systems.
    (In Ref.~\cite{MM24}, the inter-particle interaction
    constant was defined as $m^2\mathcal{C}$ because $\rho$
    is there the particle-number density.)
The constant $\mathcal{C}$, associated with the
nuclear Skyrme and molecular van der
Waals (vdW-Skyrme) interaction \cite{BB03,brguehak}, is related mainly to the
diffuse NS surface thickness $a$, Eq.~(\ref{units}).
For simplicity, one can neglect the 
correction corresponding to
the spin-orbit term known well from 
the nuclear physics; see e.g., 
Refs.~\cite{MM24,ringshuk}.
We neglect here also
    a relatively small  surface gradient correction
to the kinetic energy which is important for a
gas system different 
from that of the desired dense NS fluid
\cite{MM24}.

We should add the constraints for the variational procedure to
get equations
for the static equilibrium.
Introducing the chemical potential $\mu$ as the Lagrange
multiplier related to the particle number constraint,  
one obtains from Eqs.~(\ref{energytot}), (\ref{enerden})
and (\ref{surf}) (see Ref.~\cite{MM24}),
\be\l{eq}
\hspace{-0.5cm}\frac{\delta \epsi}{\delta \rho}\equiv
\frac{\partial\mathcal{A}}{\partial \rho}
-2 \mathcal{C}\Delta \rho =\frac{\mu}{m}~,
\ee
where $\mathcal{A}(\rho)$
  is  given by Eq.~(\ref{vol}).
 The particle number constraint
determines the chemical potential $\mu$ in terms of the
total NS particle number $N$.  
For nuclear
liquid drop, one has two equations related to the two
constraints for
the
fixed
neutron and proton numbers
 (Refs.~\cite{MS09,BM13,BM15}). 
They determine two (neutron and
proton) chemical potentials.
The Coulomb interaction can be taken into account for a proton
part of the
nucleus through the Coulomb potential; see Refs.~\cite{strtyap,MS09}.
 Equation (\ref{enerden}) for  $\mathcal{E} $
 overlaps most of the vdW-Skyrme forces \cite{CB97,vdW,CB98}.
The symmetry energy, and surface 
terms
 can be also taken into account in nuclear astrophysics
 similarly as in nuclear physics.
We should emphasize also
that there is, in fact, no strict
separation of the nuclear and gravitational contributions into the
volume  
and surface energies
because of the $\rho$  dependence of the gravitational 
energy density $\mathcal{U}(\rho)$, Eq.~(\ref{phiexp}),
and of the Lagrange equation (\ref{eq})
 for  $\rho$.

\section{Adiabatic approach}
\l{SFA}

\subsection{Macroscopic NS moments of inertia}
\l{NSMI}

For small frequencies in the statistically
averaged NS rotation description within the LPT, 
one would expect an adiabatic 
     motion. It satisfies,
in our ESA, the condition of smallness
of the rotation energy, $E_{\rm rot}$, with 
respect to the ETF energy,
$E$, Eq.~(\ref{energytot}),
\be\l{adcond}
E_{\rm rot}= \frac{1}{2}\Theta \omega^2 \ll E~.
\ee
The MI $\Theta$, Eq.~(\ref{MI}), can be evaluated in terms of 
the
energy density $\epsi(\rho)$, Eq.~(\ref{enerden}), by using the
statistically averaged component
$\tilde{\Theta}$, Eq.~(\ref{MIt}), and
its non-linear $t,\varphi$ correction 
$\mathcal{T}_{t\varphi}$, Eq.~(\ref{MItf}).
For this purpose, in the 
    linear-over-$\overline{\omega}$ approximation
for the angular momentum
$I$, 
i.e., zero approximation for the MI $\Theta$,
we neglected
all quadratic
terms over
$\overline{\omega}^2\propto \Omega^2$; see Eq.~(\ref{rotpar}).
Splitting now $\epsi(\rho)$, Eq.~(\ref{enerden}),
into the non-gradient, $\mathcal{A}(\rho)$,
Eq.~(\ref{vol}),
and gradient,
$\mathcal{B}(\nabla \rho)^2\approx \mathcal{C}
\left(\partial \rho/\partial r\right)^2$,  Eq.~(\ref{surf}),
dependent components,
one can apply similar technique as  that in 
Appendixes 
\ref{appA} and \ref{appD}, and Ref.~\cite{MM24}.
 Finally,  one can present the statistically averaged
 $\tilde{\Theta}$ and correlation $\mathcal{T}_{t\varphi}$
 MI contributions
 in terms of
 the volume, $\tilde{\Theta}_V$ and $\mathcal{T}_V$,
     and surface,
$\tilde{\Theta}_S$ and  $\mathcal{T}_S$, terms, respectively,
\be\l{adMIVS}
\tilde{\Theta}=\tilde{\Theta}_V + \tilde{\Theta}_S~, \quad
\mathcal{T}_{t\varphi}=\mathcal{T}_V + \mathcal{T}_S~.
\ee

For the statistically
averaged volume part of the MI, one finds
\be\l{adMItV}
\tilde{\Theta}_V =\frac{8\pi}{3 \bf c^2}\overline{\epsi}
\int_0^R e^{-\nu} \mathcal{J}(r)r^4\d r
=\Theta_{\rm sph}W_1\left(z^{}_0\right)~,\qquad
z^{}_0=\sqrt{1-\xi^2},\quad \xi=\frac{R}{R_{\rm S}}~, 
\ee
where $\Theta_{\rm sph}=(2/5)MR^2$ is the MI of the uniform sphere of the
NS mass $M$ and of the radius $R$,
$\overline{\epsi}$ is given by Eq.~(\ref{RS}), and
$\mathcal{J}(r)$  is the inner ($r\leq R$) part of Eq.~(\ref{Jin}).
In Eq.~(\ref{adMItV}),
\be\l{I1tV}
W_1(z^{}_0)
=\frac{20~\mathcal{I}_1(z^{}_0)}{\left(1-z^{2}_0\right)^{5/2}}~,
\ee
where
\be\l{intI1}
\mathcal{I}_1\left(z^{}_0\right)=
\int_{z^{}_0}^1 \d z\frac{(1-z^2)^{3/2}}{(2A-z)^2}
=q^{}_1(1,A)\!-\!q^{}_1\left(z^{}_0,A\right),\qquad
z=\sqrt{1-r^2/R_{\rm S}^2}~.
\ee
The transformation  of the radial integration variable
$r$ to the dimensionless one  $z$
and its upper limit $z^{}_0$ was used in these derivations.
Here,
\be\l{q1}
q^{}_1(z,A)=\frac{(z^2+6Az-24 A^2+2)\sqrt{1-z^2}}{2(2A-z)}
+\frac32 \left(8A^2-1\right)\arcsin(z)
+6A\sqrt{1-4A^2}~\ln\left[
  \zeta(z,A)\right]~,
\ee
where
\be\l{arglog}
\zeta(z,A)=
\frac{1-2 A z+\sqrt{1-4A^2}\sqrt{1-z^2}}{2A-z}~,
\ee
and $A$  is the inner
Schwarzschild metric constant
given by Eq.~(\ref{Tolconsschwm}), $A=3 z^{}_0/2$.

For the 
volume $t,\varphi$ correlation
contribution $\MItf_V$, Eq.~(\ref{MItf}), one obtains
\be 
\MItf_V 
=\frac{16\pi}{3M} \overline{\epsi}
\int_0^R e^{-\nu} \mathcal{J}(r)\tau r^2\d r 
\l{adMItfV}~,
\ee
In these calculations we
used the expressions, Eq.~(\ref{14tauinexpsol}),  for $\tau$ and its 
expansion,
Eq.~(\ref{14tauinexpsolexp4}), for analytical derivations, and 
Eq.~(\ref{RS}) with (\ref{rhobMR}),     %
\be\l{rhobMR}
\overline{\rho}=3M/(4\pi R^3)~,
\ee
for the volume energy density
$\overline{\epsi}$.  
Equation~(\ref{adMItfV}) can be conveniently presented as 
\be\l{I2tV}
\MItf_V \equiv W_2\left(\xi\right)= \frac{16 c^{}_1}{\xi^{3}}
\mathcal{I}_2\left(\xi\right) 
\ee
with the coordinate transformations from $r$ to
$x=r/R_{\rm S}$,
\be\l{IntI2}
\mathcal{I}_2= 
\int_0^{\xi} \frac{\sqrt{x}J_p\left(\sqrt{5} x\right)
  x^2\d x}{\sqrt{1-x^2}\left(2A-\sqrt{1-x^2}\right)^2}~,
\ee
and the upper integration limit $\xi$, Eq.~(\ref{adMItV}).
Using the expansion, Eq.~(\ref{14tauinexpsolexp4}),
for the Bessel function $J_p\left(\sqrt{5} x\right)$ of the order 
$p$ [Eq.~(\ref{p})] and $\sqrt{1-x^2}$ in the integrand
[Eq.~(\ref{IntI2})] over $x$, one can analytically take it
in terms of the hypergeometric (Gauss) functions $F(a,b; c; w)$;
see Eq.~(\ref{q2expJp}).
Expanding this result over $\xi$
with accounting
    for the functions $p(\xi)$ and $A(\xi)$, Eqs.~(\ref{p}) and
(\ref{Tolconsschwm}),
up to $\xi^7$ terms, one finds
\bea\l{q2exp}
\mathcal{I}^{}_2&=&\frac{\xi^5}{6\sqrt{2 \pi}~5^{1/4}}
+\frac{\xi^7}{4200 \sqrt{\pi}}
\left(
    94 \sqrt{2}~5^{3/4}+70 \sqrt{2}~5^{3/4}
    \ln{2}-35 \sqrt{2}~5^{3/4} \ln{5}\right.\nonumber\\
&&    -\left.70 \sqrt{2}~5^{3/4} \ln{\xi}+70 \sqrt{2}~5^{3/4}
    \psi(5/2)\right)~,
\eea
where $\psi(5/2) \approx 0.70$,
    Eq.~(\ref{14tauinexpsolexp4x2R2}).

Expanding also Eq.~(\ref{as6}) for $I^{}_2$ over $\xi$ up to 8th order, one
directly obtains the asymptotic expression for $\tau$ from
Eq.~(\ref{14tauin}),
avoiding the Bessel solution
expansion,
Eq.~(\ref{14tauinexpsolexp4}),
\be\l{q2as6exp2}
\mathcal{I}^{}_2 = \frac{\xi^{11/2}}{22}\left(1 + \frac{3}{2} \xi^2
+O(\xi^4)\right).
\ee
As shown in Appendix \ref{appD}, the surface MI components,
$\tilde{\Theta}_S$ and $\MItf_S$, are derived analytically 
in terms of the surface tension
coefficient $\sigma$ (see Eqs.~(\ref{enertots}) for $\sigma$ and
(\ref{Tolconsschwm}) for $A$,
 and Ref.~\cite{MM24}), 
\be\l{adMItS}
\tilde{\Theta}_S=
\frac{4\pi}{3c^2}\frac{\sigma R^4}{\sqrt{1-R^2/R_{\rm S}^2}}
\ee
and 
\be\l{adMItfS}
\MItf_S=-\frac{8\pi\sigma}{3M c^2}\frac{R^4/R_{\rm S}^2}{1-R^2/R_{\rm S}^2}~.
\ee
 They are proportional
 to the leptodermic parameter $a/R$ through the surface tension
    coefficient $\sigma$.
As shown in 
Ref.~\cite{MM24}, and
Appendix \ref{appA},
this coefficient $\sigma$
 can be calculated
 analytically; see Eq.~(\ref{enertots}).
 \begin{figure}
   \begin{center}\includegraphics[clip,width=10.0cm]{Fig2W1W2WUX_TEST.eps}
     \end{center}

   \vspace{-0.8cm}
 \caption{{\small
     Volume contributions to the moment of inertia $W_1$,
     Eq.~(\ref{I1tV}) (green dotted), and
     $\mathcal{T}_{V}\equiv W_2(\xi)$,  Eq.~(\ref{I2tV}) for $W_2$ (red dashed),
     and full volume numerical  ``1'' MI
     $W_1/(1-W_2)$ [see also Eq.~(\ref{ThetaV})];
     and its analytical Gauss  hypergeometric
     function approach ``1a''  [see red dotted, Eqs.~(\ref{I1tV}) and
       (\ref{q2expJp})] as 
     functions of the dimensionless variable, $R/R_{\rm S}$.
     They compared with the results of the
     expansion ``1b'' [blue dash-dotted, Eq.~(\ref{q2exp})] and
     the direct analytical expansion ``1c'' [thin solid magenta,
       Eq.~(\ref{q2as6exp2})].
     The coefficient $c^{}_1$ is given by Eq.~(\ref{c1}).
     Heavy dashed-cyan line displays the
     Hogan approach, Ref.~\cite{PH76}, for MI $W_1/(1-W_2)$;
     see Ref.~\cite{AM25npae} and Appendix \ref{appC}.
        The asymptotes are shown by the vertical lines.
  }}
\label{fig2}
\end{figure}

 Figure \ref{fig2} shows the volume components of the MI
which are determined by the statistically 
averaged $\tilde{\Theta}_V\propto W_1$
and 
the rotational correlation function $\mathcal{T}^{(V)}_{t,\varphi}\equiv
\mathcal{T}_{V} \equiv W_2$ [see
   Eq.~(\ref{MItf}) with $\tau$ of Eq.~(\ref{14tauinexpsol})
   for $c^{}_1$ given by Eq.~(\ref{c1})].
 MI contributions are shown as functions 
of the single dimensionless variable, 
$\xi=R/R_{\rm S}$. We introduced here the volume MI, formally 
    neglecting the
surface gradient components,
       \be\l{ThetaV}
            \Theta_V=
       \frac{\tilde{\Theta}_V}{1 - \mathcal{T}_{V}}~.
            \ee
            The results of using the analytical Gauss ``1a''
            [see Eq.~(\ref{q2expJp}) for $\mathcal{I}_2$],
            and more exact numerical Bessel ``1''
solutions, Eqs.~(\ref{I2tV}) and (\ref{IntI2}),
their leading expansion "1b" up to $x^7$ terms, Eq.~(\ref{q2exp}); and
the ``direct asymptotic $\tau$ solution ``1c'', Eq.~(\ref{q2as6exp2})
to the exact GRT equation
(\ref{14tauin}) [all
for the
full volume MI $\Theta_V\propto W_1/(1-W_2)$] are plotted too.
 Vertical lines show the asymptotes at 
 the roots, 
 $\xi=0.66$
 (black solid and very close red dotted), 
and $\xi=0.76$
 (blue dash-dotted),
 as well,  $\xi=0.80$ (solid magenta) vertical lines
to the equation, $W_2(\xi)-1=0$,
with Eqs.~(\ref{I2tV}), (\ref{IntI2}),
(\ref{q2exp}),
and (\ref{q2as6exp2}) for
$I^{}_2$, respectively. These roots at
$\xi=\xi^{}_K$, which appear because of Kerr $t,\varphi$ coupling,
           lead to a more strong rotational
constraint
for the NS radius, $\xi < \xi^{}_K$, or $R < R_{\rm K}=\xi^{}_K R_{\rm S}$. 
Here, $R_{\rm K}$ is defined, generally speaking, as a pole in Eq.~(\ref{MI})
including the surface correction, i.e.,
the root of Eq.~(\ref{RKpole}).
As seen from Fig.~\ref{fig2}, one can see 
a dramatic behavior of the black solid, red dotted,
and
solid magenta curves
near these specific points because of a strong gravity.
Notice that the analytical Gauss result,
Eq.~(\ref{q2expJp}), agrees perfectly with the numerical more exact
evaluation  of the
integral $I_2$, Eq.~(\ref{IntI2}).
The
volume component
$\tilde{\Theta}_V$ (green dotted) of the statistically averaged MI
$\tilde{\Theta}$ and the correlation contribution
$\MItf_{V}$ (red dashed lines) into $\MItf_{t\varphi}$
are mainly smooth functions of $\xi$, except for a cusp 
at $\xi=0.87$ 
in the range close to one.
For $\xi\rightarrow 1$  ($0<\xi<1$),  one has the usual 
asymptotically singular  approaching of the effective radius $R$
to the Schwarzschild radius $R_{\rm S}$, Eq.~(\ref{RS}).
We will come back to these 
remarkable properties
in the discussion Section \ref{discres}. Notice also that
the approximation ``1c''
is largely close to both the numerical
``1'' and analytical ``1a'' results and the approximation ``1b''
up to about $\xi=0.5$; as well as to the results (Appendix \ref{appC}) based on
Hogan's $\tau$, Eq.~(\ref{tauin}).
The new results for $\tau$, Eqs.~(\ref{IntI2}), (\ref{q2expJp}) and their
asymptotical expressions, Eqs.~(\ref{q2exp}),
and (\ref{q2as6exp2}),
differ quantitatively in magnitude (position of the asymptotes)
from Hogan's $\tau$ calculations for larger $\xi$;
cf. the heavy dashed cyan line [Eq.~(\ref{I2tVH})]
with our original solid black
    [Eq.~(\ref{I2tV})],
and approximate dash-dotted blue and
solid magenta results.
However, they  have a similar qualitative behavior,
and are the close the larger distance from 
the poles. 
All these compared curves converge to the same
statistically averaged (green dotted) contribution $W_1$ ``3'' at small
values of $R/R_{\rm S}$~.

 Notice that $\tilde{\Theta}_S>0$ while
$\mathcal{T}_S > 0$ because the tension coefficient $\sigma$
is positive and $R<R_{\rm K}<R_{\rm S}$ 
    for a stable equilibrium.
   As these surface components 
      are proportional
      to the leptodermic parameter $a/R$ 
          through the tension coefficient
          $\sigma$, they
              depend on the total nuclear-gravitational
          inter-particle interaction
          $\mathcal{C}$ and incompressibility
          $K_G$ through Eq.~(\ref{units}).

\subsection{Macroscopic NS masses}
\l{NSM}

For the MI calculations and 
the estimates
for the condition of
small 
frequencies $\overline{\omega}$, Eq.~(\ref{rotpar}), one
has to 
derive also the
NS masses and energies (Appendices \ref{appA} and \ref{appD}).
 Using the baryon approximation, Eq.~(\ref{RS})
 for $\overline{\epsi}$, one obtains
\be\l{MNStot}
M=\int \rho \d \mathcal{V} \approx  M_{V} + M_{S}~.
\ee
In this equation, the first volume component, $M_{V}$, 
of the NS mass is given by  
\be\l{MNSV}
M_{V}=4\pi \overline{\rho} \int_0^R
\frac{r^2{\rm d} r}{\left(1-r^2/R_{\rm S}^2\right)^{1/2}}
=
2\pi \overline{\rho} R_{\rm S}^3f\left(R/R_{\rm S}\right)~,
\ee
where
\be\l{mcor}
f(\xi)=\arcsin(\xi) - \xi \sqrt{1-\xi^2},\qquad 0 < \xi < 1~.
\ee
The second term, $M_{S}$, in Eq.~(\ref{MNStot}) is the
surface component of the NS mass, 
\be\l{MNSSgen}
M_{S}=\int \left(\rho-\overline{\rho}\right)\d V 
= 4\pi R^2a \mathcal{J}(R) \overline{\rho}
     \int_{-\infty}^{\infty}[1-y(x)] \d x~
\approx-aS\overline{\rho}\mathcal{J}(R)\int_0^1\frac{(1-y)}{
  \sqrt{\epsilon(y)}}\d y~,
\ee
where $J(R)$ is the inner 
volume-shortness factor $\mathcal{J}(r)$, 
Eq.~(\ref{Jin}), taken at $r=R$, $x=(r-R)/a$. The
integration boundaries over $x$
correspond to
the change of the density $y(x)$ from 1 to 0~.
In particular, 
in the quadratic approximation for $\epsilon(y)$,
Eq.~(\ref{varepsilon}), and the vdW-Skyrme interaction
 [see Eq.~(\ref{b0g0}) for $y(x)$],
one analytically
finds from Eq.~(\ref{MNSSgen}) 
\be\l{MNSSfin}
M_{S}\approx -8 \pi 
\overline{\rho} R^2 a \left(1-R^2/R^2_{S}\right)^{-1/2}. 
\ee
For the surface component $M_{S}$, Eq.~(\ref{MNSSfin}), in units of the
volume part $M_{V}$, Eq.~(\ref{MNSV}), one has
\be\l{MSMV}
\frac{M_{S}}{M_{V}}=
-\frac{4a R^2}{R^3_{S}f\left(R/R_{\rm S}\right)
\left(1-R^2/R^2_{\rm S}\right)^{1/2}}~.
\ee
    For understanding the contribution
        of the ES 
        into the
        leading order approximation over
        $a/R$ for a strong gravitation,
        one should compare the result
        of Eq.~(\ref{MSMV}) 
        for $M_S/M_V$ with that for a weak gravity:   
\be\l{MSMVas}
\frac{M_S}{M_V}
=6\frac{a}{R}
\left\{1+O\left[\left(\frac{R}{R_{\rm S}}\right)^5\right]\right\}~.
\ee
    We used the expansion of the volume
    gravitational factor,
    $f(\xi)=2\xi^3/3 + \xi^5/5 +O(\xi^7)$ over $\xi=R/R_{\rm S}$, and
    expansion of the
        volume-shortness factor, $\mathcal{J}(r)=\mathcal{J}(R) +
    O[(r-R)^2]$
[Eq.~(\ref{Jin})] at first order over $a/R$.
For the typical
leptodermic parameter values,
one finds from Eq.~(\ref{MSMV}) a relatively large surface mass
contribution which is proportional
to $a/R$.  In the leading leptodermic approximation
(linear over $a/R$) we neglect quadratic terms of the order of
$(a/R)^2$.
For the radial variable $r$ close to the NS radius $R$ which is,
in turn, close to the
Schwarzschild radius $R_{\rm S}$, one has
a sharp behavior of the gravitational factor $\mathcal{J}(r)$,
Eq.~(\ref{Jin}).
According to Appendix \ref{appD},
in the derivations of Eq.~(\ref{MNSSfin})
for $M_S$ [see Eq.~(\ref{Uint1})] we assume that
$\mathcal{J}(r)$ is a smooth function of the argument
$r$ 
taking off
$\mathcal{J}(r)$ from the integral, Eq.~(\ref{Uint}) at $r=R$,
to arrive at Eqs.~(\ref{Uint1}) and
(\ref{MSMV}). Therefore,
this simplest approach is valid for $\xi=R/R_{\rm S} $ values which
are enough far away
from $\xi=1$ ($\xi<1$).
In this case,  
    one has a good accuracy so that
we can neglect the
second order correction $\sim
(a/R)^2\ll 1 $~.
Largely, it means that the required accuracy should be carried out
for $\xi$ relating to the
$M_S/M_V $  up to a maximally accessible value of  $\xi$.
 \begin{table}
   \begin{center} 
     \begin{tabular}{ccccc}
\toprule 
NS & $M$ & $R$ & P & Refs.
\\
&  ($M_{\odot}$) & (km) & (ms) &   
\\
\hline
J0030+0451&\makecell{
  \\
  $1.43^{+0.20}_{-0.17}$\\
  \\
  $1.34^{+0.15}_{-0.16}$\\
}&\makecell{
  $12.68^{+1.31}_{-1.04}$\\
  \\
  $12.71^{+1.14}_{-1.19}$\\
}&\makecell{
  $4.9$\\
}&\makecell{
  \cite{YK26}\\
  \\
  \cite{TR19}\\
   }
\\
\hline
\\
J0740+6620&\makecell{
  $2.072^{+0.067}_{-0.066}$\\
  \\
  $2.08^{+0.07}_{-0.07}$\\
}&\makecell{
  $12.39^{+1.30}_{-0.98}$\\
  \\
  $12.76^{+1.49}_{-1.02}$\\
}&\makecell{
  $2.9$\\
}&\makecell{
  $\cite{TR21}$\\
  \\
  $\cite{AD24}$\\
  }
\\
\hline
\\  
J1731-347&\makecell{
  $0.77^{+0.20}_{-0.17}$\\
  \\
  $0.83^{+0.17}_{-0.13}$\\
}&\makecell{
  $10.4^{+0.86}_{-0.78}$\\
  \\
  $11.25^{+0.53}_{-0.37}$\\
}&\makecell{
  $147$\\
}&\makecell{
  $\cite{VD22}$\\
 }
\\
\hline
\\
Cen X-3&\makecell{
  $1.21^{+0.21}_{-0.21}$\\
  \\
  $1.49^{+0.08}_{-0.08}$\\
}&\makecell{
  $9.178^{+0.130}_{-0.130}$\\
}&\makecell{  
  $4840$
}&\makecell{
  $\cite{SN11}$\\
  \\
  $\cite{MR11,wiki:PSR-Cen-X-3}$\\
 }
\\
\hline
\\
4U 1702-429  &\makecell{
  $1.9^{+0.3}_{-0.3}$\\
}&\makecell{
  $12.4^{+0.3}_{-0.3}$\\
}&\makecell{
  $3.04$
}&\makecell{
  $\cite{JN17}$
  }
\\
\hline
\\
4U 1608-52 &\makecell{
  $1.74^{+0.14}_{-0.14}$\\
 }&\makecell{
  $9.3^{+1.0}_{-1.0}$\\
}&\makecell{
  $1.62$\\
}&\makecell{
  $\cite{GO10,ZR20,FO16}$\\
  }
\\
\hline
4U 1820-30  &\makecell{
  $1.58^{+0.06}_{-0.06}$\\
}&\makecell{
  $9.1^{+0.4}_{-0.4}$\\
}&\makecell{
  $1.40$
}&\makecell{
  $\cite{TG10,TG13,ZR20,FO16}$
  }
\\
\hline
\\
J1814–338  &\makecell{
  $1.21^{+0.05}_{-0.05}$\\
}&\makecell{
  $7^{+0.4}_{-0.4}$\\
}&\makecell{
  $3.18$
}&\makecell{
  $\cite{MB13,YK24}$\\
 }
\\
\hline
\\
J0437–4715 &\makecell{
  $1.418^{+0.037}_{-0.037}$\\
}&\makecell{
  $11.36^{+0.95}_{-0.63}$\\
}&\makecell{
  $5.747$
}&\makecell{
  $\cite{LM25,ZR20}$\\
 }
\\
\hline
\\
J0614–3329 &\makecell{
  $1.44^{+0.06}_{-0.07}$\\
}&\makecell{
  $10.29^{+1.01}_{-0.86}$\\
  }&\makecell{
  $3.15$
}&\makecell{
  $\cite{LM25}$\\
 }
\\
\hline
\\
J0952–0607A &\makecell{
  $2.35^{+0.17}_{-0.17}$\\
}&\makecell{
  $12.245^{+0.685}_{-0.315}$\\
}&\makecell{
  $1.41$\\
}&\makecell{
  $\cite{RK23,wiki:PSR-J0952-0607}$\\
  }
\\
\hline
\\
Vela Pulsar &\makecell{
  $1.88^{+0.13}_{-0.13}$\\
}&\makecell{
  $12.91^{+0.39}_{-0.39}$\\
}&\makecell{
  $89$\\
}&\makecell{
  $\cite{DS24}$\\
  }
\\
\hline
\\
Vela X-1 &\makecell{
  $1.77^{+0.08}_{-0.08}$\\
  \\
   $2.12^{+0.16}_{-0.16}$\\
}&\makecell{
  $9.56^{+0.08}_{-0.08}$\\
}&\makecell{
  $283\cdot 10^3$\\
}&\makecell{
  $\cite{MR11,TG13,MF15,GN23,SS23}$\\
  }
\\
\hline
\\
SAX J1748.9-2021 &\makecell{
  $1.81^{+0.25}_{-0.37}$\\
}&\makecell{
  $11.7^{+1.7}_{-1.7}$\\
}&\makecell{
  $2.26$\\
}&\makecell{
  $\cite{RK23,ZR20,FO16}$\\
  }
\\
\hline
\\
EXO 1745-268 &\makecell{
  $1.65^{+0.21}_{-0.31}$\\
}&\makecell{
  $10.5^{+1.6}_{-1.6}$\\
}&\makecell{
  1.65
}&\makecell{
  $\cite{RK23,ZR20,FO16}$\\
  }
\\
\hline
\\
KS 1731-260 &\makecell{
  $1.61^{+0.35}_{-0.37}$\\
}&\makecell{
  $10.2^{+2.2}_{-2.2}$\\
}&\makecell{
  $1.91$\\
}&\makecell{
  $\cite{RK23,ZR20,FO16}$\\
  }
\\
\hline
\\
     \end{tabular}
   \end{center}  

\vspace{-0.6cm}
\caption{{\small
    Observational  values 
    for masses $M$ (second), radii $R$ (third),
    rotational periods $P$ (fourth), and references (fifth),
        are shown for
        the neutron stars named in the first column
            up to J0614-3329. For other
        NSs the radii $R$ are found by using the model calculations.
  }}
\label{table-1}
\end{table}

Notice also that due to 
the integration over
the 
  radial variable $r$,
  the smallness factor, proportional to $a/R$, appears for 
 the  NS surface mass $M_S$,
  Eq.~(\ref{MNSSfin}),
  as well as in $\Theta_S$ for the MI, and $E_S$,
  Eq.~(\ref{enertots}), through the tension coefficient $\sigma$
  [Eq.~(\ref{sigma1})].
   Thus, 
      these surface components, $M_{S}$ and $E_{S}$
      are relatively of the order of a small parameter $a/R$.
      However, the surface component of the local NS
      characteristics, e.g., 
      the vdW capillary surface pressure (including the
      gravitational field) equilibrates the
      volume pressure acting from the interior of a finite
          fluid system to
          be a leading reason of its equilibrium
          stability \cite{MM25npa}.
          As seen from the surface corrections
                to the mass $M$,
          Eq.~(\ref{MNSSfin}), and the MI contributions $\tilde{\Theta}$,
          Eq.~(\ref{adMItS}) and
          $\MItf$, Eq.~(\ref{adMItfS}), a strong gravity
          leads to the 
                     constraints on the NS radii $R$ within
               $r^{}_g < R < R_{\rm K}< R_{\rm S}$~.

\subsection{Adiabatic condition}
\l{adiabcond}

 \begin{table}
   \begin{center} 
     \begin{tabular}{ccccc}
\toprule 
NS &  $Ic/M^2G$& $Ic/M^2G$& $P_0$
& $P_0$
\\
&  \cite{PH76}&  & \cite{PH76}& (ms) 
\\
\hline
J0030+0451&\makecell{
  $0.39~;~0.46$\\
  $0.39~;~0.50$\\
  \\
  $0.38~;~0.47$\\
  $0.39~;~0.51$\\
 }&\makecell{
  $0.50~;~0.53$\\
  $0.44~;~0.54$\\
  \\
  $0.46~;~0.52$\\
  $0.44~;~0.55$\\
}&\makecell{
  $0.14~;~0.15$\\
  $0.13~;~0.15$\\
  \\
  $0.13~;~0.15$\\
  $0.12~;~0.14$\\
}&\makecell{
  $0.13~;~0.15$\\
  $0.14~;~0.15$\\
  \\
  $0.15~;~0.16$\\
  $0.13~;~0.15$\\
}
\\
\hline
J0740+6620&\makecell{
  $0.99~;~0.82$\\
  $0.81~;~0.78$\\
  \\  
  $0.94~;~0.83$\\
  $0.79~;~0.80$\\
}&\makecell{  
  $5.34~;~1.23$\\
  $1.83~;~1.07$\\
  \\
  $3.10~;~1.18$\\
  $1.55~;~1.05$\\
}&\makecell{
  $0.19~;~0.18$\\
  $0.17~;~0.17$\\
  \\
  $0.18~;~0.18$\\
  $0.17~;~0.17$\\
}&\makecell{
  $0.44~;~0.22$\\
  $0.25~;~0.20$\\
  \\
  $0.34~;~0.21$\\
  $0.23~;~0.20$\\
}
\\
\hline
J1731-347&\makecell{
  $1.10~;~1.38$\\
  $1.44~;~1.88$\\
  \\
  $1.30~;~1.46$\\
  $1.59~;~1.82$\\
}&\makecell{
  $1.21~;~1.47$\\
  $1.48~;~1.92$\\
  \\
  $1.40~;~1.55$\\
  $1.64~;~1.87$\\
}&\makecell{
  $0.10~;~0.12$\\
  $0.10~;~0.11$\\
  \\
  $0.11~;~0.12$\\
  $0.11~;~0.12$\\
}&\makecell{
  $0.11~;~0.12$\\
  $0.10~;~0.11$\\
  \\
  $0.12~;~0.13$\\
  $0.11~;~0.12$\\
}
\\
&\makecell{
  $\times 10^{-2}$\\
  }&\makecell{
  $\times 10^{-2}$\\
}&\makecell{ 
  }
  \\
&\makecell{
  }
  \\
\hline 
Cen X-3&\makecell{
  $3.25~;~3.26$\\
  $3.06~;~3.17$\\
  \\
   $3.74~;~3.65$\\
  $3.23~;~3.25$\\
}&\makecell{
  $4.93~;~4.72$\\
  $3.47~;~3.56$\\
  \\
  $7.82~;~6.73$\\
  $4.84~;~4.65$\\
}&\makecell{
  $0.12~;~0.12$\\
  $0.10~;~0.10$\\
  \\
  $0.13~;~0.13$\\
  $0.12~;~0.12$\\
}&\makecell{
  $0.14~;~0.14$\\
  $0.11~;~0.11$\\
  \\
  $0.19~;~0.18$\\
  $0.14~;~0.14$\\
}\\
&\makecell{
  $\times 10^{-4}$\\
  }&\makecell{
  $\times 10^{-4}$\\
}&\makecell{ 
}\\
&\makecell{
}\\
\hline
4U 1702-429  &\makecell{
  $0.95~;~0.80$\\
  $0.63~;~0.72$\\
}&\makecell{
  $3.97~;~1.31$\\
  $0.79~;~0.84$\\
}&\makecell{
  $0.19~;~0.18$\\
  $0.14~;~0.15$\\
}&\makecell{ 
  $0.39~;~0.23$\\
  $0.16~;~0.16$\\
 }\\
\hline
4U 1608-52  &\makecell{
  $-1.65~;~1.45$\\
  $~1.49~;~1.09$\\
}&\makecell{
  $-0.63~~;~4.60$\\
  $-135.8~;~1.63$\\
 }&\makecell{
  $0.16~;~0.16$\\
  $0.15~;~0.13$\\
}&\makecell{ 
  $0.10~;~0.29$\\
  $1.41~;~0.16$\\
 }\\
\hline
4U 1820-30  &\makecell{
  $1.39~;~1.16$\\
  $1.09~;~1.04$\\
}&\makecell{
  $9.08~;~2.36$\\
  $2.37~;~1.64$\\
}&\makecell{
  $0.15~;~0.14$\\
  $0.13~;~0.12$\\
}&\makecell{ 
  $0.37~;~0.19$\\
  $0.19~;~0.16$\\
}\\
\hline
J1814–338  &\makecell{
  $0.57~;~0.45$\\
  $0.43~;~0.40$\\
}&\makecell{
  $8.49~;~0.86$\\
  $0.96~;~0.61$\\
}&\makecell{
  $0.11~;~0.10$\\
  $0.10~;~0.10$\\
}&\makecell{  
  $0.44~;~0.14$\\
  $0.15~;~0.12$\\
 }\\
\hline
J0437–4715  &\makecell{
  $0.30~;~0.35$\\
  $0.30~;~0.35$\\
}&\makecell{
  $0.38~;~0.40$\\
  $0.37~;~0.40$\\
}&\makecell{
  $0.13~;~0.14$\\
  $0.12~;~0.14$\\
}&\makecell{  
  $0.14~;~0.15$\\
  $0.14~;~0.15$\\
 }\\
\hline
J0614–3329 &\makecell{
  $0.53~;~0.58$\\
  $0.49~;~0.58$\\
}&\makecell{
  $0.82~;~0.72$\\
  $0.67~;~0.68$\\
}&\makecell{
  $0.12~;~0.13$\\
  $0.12~;~0.13$\\
}&\makecell{  
  $0.15~;~0.15$\\
  $0.14~;~0.14$\\
 }\\
\hline
J0952–0607A &\makecell{
  $1.02~;~2.87$\\
  $1.94~;~1.74$\\
}&\makecell{
  $-(2.35~;~24.51)$\\
  $~6.24~;~3.20$\\
 }&\makecell{ 
  $0.46~;~0.24$\\
  $0.19~;~0.18$\\
 }&\makecell{  
  $0.21~;~0.70$\\
  $0.33~;~0.24$\\
}\\
\hline
Vela Pulsar &\makecell{
  $2.49~;~2.51$\\
  $2.28~;~2.39$\\
}&\makecell{
  $3.97~;~3.58$\\
  $2.96~;~2.96$\\
 }&\makecell{ 
  $0.17~;~0.17$\\
  $0.15~;~0.16$\\
 }&\makecell{  
  $0.21~;~0.20$\\
  $0.17~;~0.17$\\
}\\
&\makecell{
  $\times 10^{-2}$\\
  }&\makecell{
  $\times 10^{-2}$\\
}&\makecell{ 
}\\
&\makecell{
}\\
\hline 
Vela X-1&\makecell{
  $1.06~;~0.97$\\
  $0.71~;~0.70$\\
  \\
   $-(0.34~;~0.46)$\\
  $~2.13~;~1.62$\\
}&\makecell{
  $-8.07~;~36.10$\\
  $1.78~;~1.54$\\
  \\
  $-(0.22~;~0.27)$\\
  $-(1.30~;1.81)$\\
}&\makecell{
  $0.17~;~0.17$\\
  $0.14~;~0.14$\\
  \\
  $0.10~;~0.12$\\
  $0.25~;~0.22$\\
}&\makecell{
  $0.49~;~1.03$\\
  $0.22~;~0.21$\\
  \\
  $0.08~;~0.09$\\
  $0.20~;~0.23$\\
}\\
&\makecell{
  $\times 10^{-5}$\\
  }&\makecell{
  $\times 10^{-5}$\\
}&\makecell{ 
}\\
&\makecell{
}\\
\hline
SAX J1748.9-2021  &\makecell{
  $2.62~;~1.01$\\
  $0.73~;~0.98$\\
}&\makecell{
  $-1.83~;~1.48$\\
  $~0.97~;~1.10$\\
}&\makecell{
  $0.25~;~0.17$\\
  $0.12~;~0.15$\\
}&\makecell{ 
  $0.22~;~0.21$\\
  $0.14~;~0.16$\\
}\\
\hline
EXO J1745-268  &\makecell{
  $4.54~;~1.25$\\
  $0.91~;~1.20$\\
}&\makecell{
  $-1.75~;~1.83$\\
  $~1.29~;~1.36$\\
 }&\makecell{
  $0.27~;~0.15$\\
  $0.11~;~0.13$\\
}&\makecell{ 
  $0.13~;~0.19$\\
  $0.17~;~0.14$\\
 }\\
\hline
KS 1731-260  &\makecell{
  $-3.01~;~1.13$\\
  $~0.72~;~1.10$\\
}&\makecell{
  $-0.22~;~1.75$\\
  $~1.07~;~1.21$\\
}&\makecell{
  $0.07~;~0.16$\\
  $0.10~;~0.13$\\
}&\makecell{ 
  $0.06~;~0.20$\\
  $0.13~;~0.14$\\
}\\
\hline
\\
     \end{tabular}
    \end{center} 

\vspace{-0.8cm}
\caption{{\small
      Angular momenta $Ic/M^2G$ for
    the Hogan value of $\tau$, Eq.~(\ref{tauin}) \cite{PH76} (second column),
    and 
    that for our derivations of $\tau$,
    Eq.~(\ref{14tauinexpsol})
    (third column)
    are shown for
    the NSs named in first columns in
    Tables \ref{table-1}-\ref{table-3}.
   The full
      limit (absolute value)
   periods $P_0$ 
   [Eq.~(\ref{omcond})], 
    are displayed in the fourth and fifth columns, respectively.
     The incompressibility parameter
    $\kappa=10$ for a strong gravity [Eq.~(\ref{kappa})],  
       and a leptodermic 
   parameter $a/R=0.08$ are used.
            Lower and upper rows 
        in
        each NS line 
        are related to the
        minimal and maximal values of the NS mass $M$, according to the
        errors
        in the second column of Table \ref{table-1}.
        Minimal (left) and maximal (right) values of the NS radii $R$,
        separated by
        a semicolon, correspond to their mean values with
        errors shown in third column of Table \ref{table-1}.
   }}
\label{table-2}
 \end{table}
  \begin{table}
   \begin{center} 
     \begin{tabular}{ccccc}
\toprule 
NS & $Ic/M^2G$& $Ic/M^2G$
& $Ic/M^2G$& $R/R_{\rm S}$
\\
& (a) & (b)  & (c) &  
\\
\hline
J0030+0451&\makecell{
$0.50~;~0.53$\\
  $0.44~;~0.59$\\
  \\
$0.46~;~0.57$\\
  $0.43~;~0.61$\\
  }&\makecell{
$0.24~;~0.37$\\
  $0.29~;~0.45$\\
  \\
$0.25~;~0.40$\\
  $0.30~;~0.48$\\
  }&\makecell{
$0.29~;~0.43$\\
  $0.33~;~0.50$\\
  \\
$0.30~;~0.45$\\
  $0.34~;~0.53$\\
  }&\makecell{
  $0.64~;~0.59$\\
  $0.57~;~0.52$\\
  \\
$0.62~;~0.56$\\
  $0.55~;~0.50$\\
  }
\\
\hline
J0740+6620&\makecell{
  $3.61~;~1.10$\\
  $1.64~;~1.03$\\
  \\
  $2.51~;~1.11$\\
  $1.43~;~1.03$\\
  }&\makecell{
  $0.33~;~0.51$\\
  $0.34~;~054$\\
  \\
  $0.34~;~0.51$\\
  $0.36~;~0.54$\\
}&\makecell{
  $0.39~;~0.62$\\
  $0.41~;~0.64$\\
  \\
  $0.41~;~0.62$\\
  $0.43~;~0.64$\\
}&\makecell{
  $0.74~;~0.68$\\
  $0.72~;~0.66$\\
  \\
  $0.74~;~0.67$\\
  $0.71~;~0.65$\\
  }
\\
\hline
J1731-347&\makecell{
  $1.21~;~2.25$\\
  $1.48~;~3.16$\\
  \\
  $1.39~;~2.22$\\
  $1.64~;~2.81$\\
}&\makecell{
  $0.84~;~1.88$\\
  $1.24~;~2.87$\\
  \\
  $1.02~;~1.83$\\
  $1.37~;~2.50$\\
}&\makecell{
  $0.96~;~2.02$\\
  $1.33~;~2.98$\\
  \\
  $1.15~;~1.98$\\
  $1.47~;~2.62$\\
}&\makecell{
  $0.55~;~0.50$\\
  $0.43~;~0.40$\\
  \\
  $0.52~;~0.50$\\
  $0.44~;~0.42$\\
}\\
&\makecell{
  $\times 10^{-2}$\\
  }&\makecell{
  $\times 10^{-2}$\\
}&\makecell{ 
  $\times 10^{-2}$\\
}&\makecell{
}\\
\hline
Cen X-3&\makecell{
  $4.76~;~5.79$\\
  $3.45~;~6.73$\\
  \\
  $7.12~;~5.71$\\
  $4.67~;~5.80$\\
}&\makecell{
  $1.74~;~4.17$\\
  $2.23 ~;~5.56$\\
  \\
  $1.63~;~3.86$\\
  $1.75~;~4.19$\\
}&\makecell{
  $2.10~;~4.72$\\
  $2.59~;~6.01$\\
  \\
  $1.96~;~4.43$\\
  $2.11~;~4.74$\\
 }&\makecell{ 
  $0.68~;~0.67$\\
  $0.57~;~0.56$\\
  \\
  $0.72~;~0.71$\\
  $0.68~;~0.67$\\
}\\
&\makecell{
  $\times 10^{-4}$\\
  }&\makecell{
  $\times 10^{-4}$\\
}&\makecell{ 
  $\times 10^{-4}$\\
}&\makecell{
}\\
\hline
4U 1702-429  &\makecell{
  $2.96~;~1.09$\\
  $0.78~;~0.91$\\
}&\makecell{
  $0.33~;~0.48$\\
  $0.40~;~0.61$\\
}&\makecell{
  $0.39~;~0.58$\\
  $0.48~;~0.70$\\
}&\makecell{
  $0.74~;~0.69$\\
  $0.63~;~0.59$\\
}\\
\hline
4U 1608-52  &\makecell{
  $-7.35 ~;~1.77$\\
  $13.70~;~1.71$\\
}&\makecell{
  $0.38~;~1.01$\\
  $0.42~;~1.14$\\
}&\makecell{
  $0.40~;~1.12$\\
  $0.49~;~1.31$\\
}&\makecell{
  $0.82~;~0.72$\\
  $0.75~;~0.66$\\
}\\
\hline
4U 1820-30  &\makecell{
  $5.62~;~1.71$\\
  $2.14~;~1.71$\\
}&\makecell{
  $0.44~;~1.12$\\
  $0.47~;~1.18$\\
}&\makecell{
  $0.52~;~1.29$\\
  $0.56~;~1.35$\\
}&\makecell{
  $0.75~;~0.44$\\
  $0.72~;~0.42$\\
}\\
\hline
J1814–338  &\makecell{
  $3.27~;~0.92$\\
  $0.86~;~0.95$\\
  }&\makecell{
  $0.17~;~0.70$\\
  $0.18~;~0.75$\\
 }&\makecell{
  $0.20~;~0.78$\\
  $0.22~;~0.82$\\
 }&\makecell{
  $0.75~;~0.71$\\
  $0.72~;~0.68$\\
}\\
\hline
J0437–4715  &\makecell{
  $0.38~;~0.49$\\
  $0.36~;~0.49$\\
  }&\makecell{
  $0.19~;~0.24$\\
  $0.20~;~0.25$\\
 }&\makecell{
  $0.23~;~0.39$\\
  $0.24~;~0.41$\\
 }&\makecell{
  $0.63~;~0.59$\\
  $0.62~;~0.58$\\
 }\\
\hline
J0614–3329 &\makecell{
  $0.79~;~0.88$\\
  $0.65~;~0.90$\\
  }&\makecell{
  $0.28~;~0.61$\\
  $0.29~;~0.66$\\
 }&\makecell{
  $0.33~;~0.70$\\
  $0.35~;~0.74$\\
 }&\makecell{
  $0.69~;~0.63$\\
  $0.65~;~0.60$\\
}
\\
\hline
J0952–0607A &\makecell{
  $-2.94~;~3.41$\\
  $~5.09~;~2.33$\\
}&\makecell{
  $0.66~;~0.95$\\
  $0.71~;~1.04$\\
  }&\makecell{
  $0.73~;~1.14$\\
  $0.85~;~1.25$\\
    }&\makecell{
  $0.79~;~0.76$\\
  $0.73~;~0.71$\\
}\\
\hline
Vela Pulsar &\makecell{
  $3.82~;~3.49$\\
  $2.92~;~2.93$\\
}&\makecell{
  $1.28~;~1.42$\\
  $1.41~;~1.56$\\
 }&\makecell{ 
  $1.55~;~1.71$\\
  $1.69~;~1.86$\\
 }&\makecell{  
  $0.69~;~0.67$\\
  $0.64~;~0.62$\\
}\\
&\makecell{
  $\times 10^{-2}$\\
  }&\makecell{
  $\times 10^{-2}$\\
}&\makecell{$\times 10^{-2}$ 
}\\
&\makecell{
}\\
\hline 
Vela X-1&\makecell{
  $~-236.8;~7.43$\\
  $1.58~;~1.40$\\
  \\
   $-(0.25~;~0.31)$\\
  $-(1.66~;~2.49)$\\
}&\makecell{
  $0.27~;~0.28$\\
  $0.29~;~0.29$\\
  \\
  $0.25~;~0.25$\\
  $0.26~;~0.27$\\
}&\makecell{
  $0.32~;~0.33$\\
  $0.34~;~0.35$\\
  \\
  $0.24~;~0.25$\\
  $0.30~;~0.31$\\
}&\makecell{
  $0.76~;~0.75$\\
  $0.73~;~0.72$\\
  \\
  $0.84~;~0.84$\\
  $0.78~;~0.78$\\
}\\
&\makecell{
  $\times 10^{-5}$\\
  }&\makecell{
  $\times 10^{-5}$\\
}&\makecell{$\times 10^{-5}$
}\\
&\makecell{
}\\
\hline
SAX J1748.98-2021  &\makecell{
  $-2.44~;~1.35$\\
  $~0.96~;~1.24$\\
}&\makecell{
  $0.35~;~0.68$\\
  $0.44~;~0.88$\\
}&\makecell{
  $0.39~;~0.81$\\
  $0.53~;~1.00$\\
}&\makecell{
  $0.78~;~0.67$\\
  $0.65~;~0.56$\\
}\\
\hline
EXO 1745-268  &\makecell{
  $-2.24~;~1.73$\\
  $~1.25~;~1.73$\\
}&\makecell{
  $0.42~;~1.00$\\
  $0.52~;~1.28$\\
}&\makecell{
  $0.47~;~1.18$\\
  $0.62~;~1.44$\\
}&\makecell{
  $0.79~;~0.67$\\
  $0.67~;~0.57$\\
}\\
\hline
KS 1731-260  &\makecell{
  $-0.26~;~1.54$\\
  $~1.03~;~1.53$\\
}&\makecell{
  $0.31~;~0.84$\\
  $0.39~;~1.18$\\
}&\makecell{
  $0.29~;~0.99$\\
  $0.47~;~1.31$\\
}&\makecell{
  $0.85~;~0.68$\\
  $0.68~;~0.54$\\
}\\
\hline
\\
\end{tabular}
\end{center}

\vspace{-0.6cm}
\caption{{\small 
        The same as in the third column of the Table \ref{table-2} but for
        different approaches: (a) for the Gauss approximation,
    Eq.~(\ref{q2expJp});
    (b) for its asymptote,
    Eq.~(\ref{q2exp}); (c) for the asymptotical expansion
    of solutions to the exact equation~(\ref{q2as6exp2})
    up to eighth
    order over $\xi$. The last column shows the key NS parameter
    $R/R_{\rm S}$. 
}}
\label{table-3}
  \end{table}
Using now Eqs.~(\ref{MI}), (\ref{adMIVS})-(\ref{adMItfS}),
one can specify
      the adiabatic condition (\ref{adcond}) for the 
       NS rotation
            periods $P=2\pi/\omega$, 
      \be\l{omcond}
           P\gg P_0=2\pi\sqrt{\frac{\Theta}{2E}}~,
           \ee
      \noindent where $P_0$ is the characteristic period limit
      for an adiabatic motion.
                Using
      the expression for the volume MI
      component, Eq.~(\ref{ThetaV}),
         one can
      largely      
      approximate the adiabatic characteristic-period limit $P_0$
      by the volume components. 
      In this volume approximation, 
      $P_0\approx P_0^{(V)}$, 
      one obtains
      \be\l{om0V1}
      P^{(V)}_0=2\pi\!\sqrt{\frac{
          \tilde{\Theta}_V}{2E_{V}\left(1\!-\!\mathcal{T}_V\right)}}
      \approx\frac{2\pi R}{c}
      \sqrt{\frac{W_1(R/R_{\rm S})}{1-W_2(R/R_{\rm S})}}.
         \ee
   see Eqs.~(\ref{adMItV}) for $\tilde{\Theta}_V$,
           (\ref{adMItfV}) for $\mathcal{T}_V$, (\ref{enertotv})
           for  $E_V$,  
            (\ref{RS}) and (\ref{rhobMR}) for $\overline{\epsi}$~.
           The characteristic statistically averaged
         periods $\tilde{P}_0^{(V)}$
         are given by Eq.~(\ref{om0V1}) as 
         \bea\l{P0Vt}
      \tilde{P}^{(V)}_0 &=&
      2\pi\sqrt{\frac{\tilde{\Theta}_V}{2E_{V}}}
      \approx \frac{2\pi R}{c}
      \sqrt{W_1\left(\frac{R}{R_{\rm S}}\right)}\\
                  &\approx& (0.11~-~0.27)~\mbox{ms}.  
                  \l{P0Vtn}
         \eea
 These values of $\tilde{P}^{(V)}_0$
      are presented
      for the two limit 
      NS masses,
      $M \approx 
      (0.6~-~2.5)M_{\odot}$ (Table \ref{table-1}),
      obtained 
      recently in the observations,
          Refs.~
         \cite{JN17,TR19,MM19,TR21,VD22,AD24},
     $M_{\odot}$ is the solar mass,
     $M_{\odot}=2.0 \cdot 10^{30}$ kg. A typical NS radius
        $R = 10$ km, 
         and leptodermic parameter,
         $a/R=0$, are used in these estimates.
         For a larger effective radius, $R = 15$ km, one finds
         $\tilde{P}_0^{(V)}\approx 
              (0.15~-~0.24)~\mbox{ms}$.
         With increasing $R/R_{\rm S}\propto \sqrt{\overline{\rho}}$, or
         density $\overline{\rho}$
         for a given radius $R$,
         the upper limit $\tilde{P}_0^{{V}}$ has a maximum.
                  A similar behavior, takes place with growing $R$
         for a given $\overline{\rho}$. 

           \section{Discussions of the results}
      \l{discres}
             \begin{figure}  
 \includegraphics[clip,width=6.9cm]{Fig3a_q1-4_k10_aR08_b0newest.eps}  
           ~
  \includegraphics[clip,width=6.9cm]{Fig3b_TRRS.eps}   

   \vspace{-0.4cm}
 \caption{{\small
        Adiabatic NS moments of inertia, $\Theta$, Eq.~(\ref{MI}),
        in units of the uniform sphere MI, 
        $\Theta_{\rm sph}=$ $2 M R^2/5$
        with the same radius
        $R$ and mass $M$, as functions of
        the effective radius $R$ 
        are shown
        for several values of 
        the inner
         densities 
        $\overline{\rho}/\rho^{}_0$=1 
        (solid), 2
         (dashed), 3
        (dotted), and 4 (dash-dotted lines)
        where $\rho^{}_0=m^{}_{N}n^{}_0$=2.68$\cdot
        10^{14} \mbox{g}/\mbox{cm}^3$ 
            is the nuclear matter density ($m^{}_{N}$ is 
        the nucleon mass, $n^{}_0=0.16$ fm$^{-3}$
        is the nuclear-matter particle-number density)
            for the incompressibility
            $\kappa=10$, Eq.~(\ref{kappa}), and leptodermic parameter
            $a/R=0.08$. 
            (a): Hogan's metric for the inner $\tau$, Eq.~(\ref{tauin}).
            (b): our metric solution for the inner $\tau$,
            Eq.~(\ref{14tauinexpsol}).  
            Vertical lines and arrows 1~-~4 show the
            asymptotes,  which are
            related to the rotational critical radius 
                 $R_{\rm K}$ corresponding to a  root of Eq.~(\ref{RKpole}).
        }}
\label{fig3}
 \end{figure}
             Table \ref{table-1} displays the mass-radius data for
             neutron stars. 
             Tables \ref{table-2} and
             \ref{table-3}, and Figures \ref{fig3}~-~\ref{fig7} 
show the results of our calculations.
In  Figs.~\ref{fig3} and \ref{fig4} we plot
      the adiabatic NS 
      moments of inertia $\Theta(R)$,
       Eqs.~(\ref{MI}) with (\ref{adMIVS}),
      in units of the uniform sphere MI,
      $\Theta_{\rm sph}$, 
           as functions
      of the effective radius $R$ 
      for several parameters. 
These MI are shown for
      different  NS values of the asymptotic
           inner
       densities 
      $\overline{\rho}/\rho^{}_0$ =1~-~ 4
      (in units of the 
      mass density of 
      nuclear matter $\rho^{}_0$), the fixed 
      leptodermic parameter
      $a/R=0.08$, and incompressibility $\kappa=10$ in 
              a strong
              gravitation case Fig.~\ref{fig3}(a) for Hogan's
              [Eq.~(\ref{tauin})] and our Fig.~\ref{fig3}(b)
              [Eq.~(\ref{14tauinexpsol})] $\tau$ functions, respectively.
       We introduced here $\kappa$ as  the relative 
      total dimensionless
    incompressibility (Ref.~\cite{MM24}),
    \be\l{kappa}
    \kappa=
    \frac{\overline{\rho} K_G}{12 m \overline{\epsi}}
    \approx \frac{K_G}{12 m c^2}~, 
    \ee
    where
    $K_G$ is the total incompressibility
        including the gravitational part
    [Eq.~(\ref{KG})], $m$ is the test-particle mass,
        $\overline{\epsi}$ is determined by Eq.~(\ref{RS}).
           For nuclear physics,
   one finds $\kappa\approx 1$ ($K_G=K\approx 240 $ MeV,
    $\overline{\epsi}=-b^{}_V n^{}_0$, $\overline{n}_0=\overline{\rho}/m^{}_N$,
    $m^{}_N$ is the nucleon mass, $b_V \approx -16$ MeV
    is the
    volume nuclear-matter separation energy,
    and $\kappa=1.25$). 
        For a weak gravitation of the order of a nuclear matter
    interaction, one can put $\kappa\approx 2$~. For
    a strong NS gravitational field, 
   we may assume $\kappa \gg 1$. Then, 
   according to Eq.~(\ref{sigmacorr}),
    one can neglect
    the correction $\sigma_{\rm corr}$
    to the tension coefficient $\sigma$.
           The NS MIs 
              in Fig.~\ref{fig4}(a)
      are considered
      at  
      a certain 
      value of the parameter $\overline{\rho}/\rho^{}_0 =3$
     for our expression
      for $\tau$,
      (and the same $\kappa=10$ and
          $a/R$=0.08), 
      but they are compared with several approaches.           
       We show a significant influence of  
      the $t,\varphi$
      correlation
      contribution $\mathcal{T}_{t\varphi}$ and its 
      volume component, $\mathcal{T}_{V}$, 
      cf. solid black 
          with and dashed red without $\mathcal{T}_{t\varphi}$
          curves in Fig.~\ref{fig4}(a).
      The surface contribution shown by 
      difference of these curves from the 
          corresponding dotted green ones
       enhances much with including this 
      correlation. 
               The Schwarzschild radius
                $R_{\rm S}$ 
      is calculated as function of the 
      density $\overline{\rho}$
      through Eq.~(\ref{RS}) for the 
      inner energy density 
      $\overline{\epsi}$.
    \begin{figure}  
     \includegraphics[clip,width=6.9cm]{Fig4a_TRRSrho3.eps}
     \hspace{0.5cm}
      \includegraphics[clip,width=6.9cm]{Fig4b_Trho.eps}
   
   \vspace{-0.4cm}         
\caption{{\small
            (a): The same as in Fig.~\ref{fig3}(b) but
        for a given value of 
        the inner 
        density,
        $\overline{\rho}/\rho^{}_0=3$, 
         is compared with different
        approximations. 
              Solid black, Eq.~(\ref{MI}),  
              and dashed red, Eq.~(\ref{MIt}) 
              [see also Eqs.~(\ref{adMItV}), (\ref{adMItS}),
             (\ref{adMItfV}) and (\ref{adMItfS})] lines show 
               the MI contributions with and without
               correlation component 
               $\mathcal{T}_{t\varphi}$, respectively. 
               The volume contributions are 
               correspondingly shown by 
             frequent and rare green dotted 
             lines; see
               Eq.~(\ref{ThetaV}).
        The arrows show  
        the
        Schwarzschild radius value $R_{\rm S}$, 
        Eq.~(\ref{RS}) 
            (right dashed red and green dotted asymptotes), 
         and the poles $R_{\rm K}$
        as the positions of asymptotes for the black solid (black arrow) and 
        frequent dotted (green one) lines.
        (b): The same as in (a) but
         as functions of
         the relative mean
         density
         $\overline{\rho}/\rho^{}_0$
         are shown
         for several values of the typical effective 
         radii, $R$=10~-~14 km, and different crust
         thickness, $a/R=0.04~-~0.16$. 
                Arrows 
                $Q_{\rm K}$ correspond to
          those of 
         $R_{\rm K}$ values
          in Fig.~\ref{fig4}(a) but in terms of
          the $\overline{\rho}/\rho^{}_0$ variable.
}}
\label{fig4}
\end{figure}
          As seen from 
          Fig.~\ref{fig4}(a), 
          the $t,\varphi$ correlation 
          contribution, $\mathcal{T}_{t\varphi}$, is 
          responsible for a dramatic behavior of the 
          curve shapes near the
              root $R_K=\xi^{}_KR_{\rm S}$ of
              Eq.~(\ref{RKpole}) (see
              discussions of Fig.~\ref{fig2} for the volume and
              at the end
              of Sect.~\ref{NSM} for the surface components),
           $R_{\rm K} < R_{\rm S}$,
          where one has a pole of the MI, Eq.~(\ref{MI}),
          $\mathcal{T}_{t\varphi}=1$. 
            The surface component, 
           $\mathcal{T}_S$, leads to the essential correction
           at smaller radii $R$ near $R_{\rm K}$ 
           due to the correlation effect, 
           in contrast to a much smaller 
           contribution of $\tilde{\Theta}_S$ everywhere 
           in the statistically averaged MI,
           $\tilde{\Theta}_S$. 
          The surface components can not be 
         considered closely to both
         asymptotes by the same reason as that
         in derivations of the
         surface tension coefficient $\sigma$ 
         (Appendix \ref{appA}) and NS surface mass $M_S$,
              Eq.~(\ref{MNSSfin}). 
         As shown in Appendix  
         \ref{appD} and Ref.~\cite{MM24},
     the derivations of the surface MI components, $\tilde{\Theta}_S$
and $\mathcal{T}_S$, 
are doubtful 
    in a small area near the asymptote, $R_{\rm K}$ (or $R_{\rm S}$),
    because the volume-shortness factor $\mathcal{J}(r)$
    and others 
 are not a smooth function of
the radial coordinate $r$ near the ES, $r=R$, in the case of
$R \rightarrow R_{\rm K}$, in particular, for $R \rightarrow R_{\rm S}$. 
We should remind that the significant   
effect with
accounting 
for the rotational correlation, $\mathcal{T}_{t\varphi}$,
is due to the gravitation self-consistent relation for 
 the MI  $\Theta$; see Eq.~(\ref{Theta2}).

     Figure \ref{fig4}(b) presents the same  
 dramatic  MI behavior as in Fig.~\ref{fig4}(a)
 but as function of the
      density $\overline{\rho}$ for different typical
      effective radii $R$, and of the leptodermic parameter $a/R$.
      Other parameters are the same as in Fig.~\ref{fig4}(a).
       Sometimes, it might be more convenient for a comparison
      with observational
      data, especially for the NSs for which the effective radii
$R$ were measured along with their masses with good accuracy
(Table \ref{table-1}). As seen from these
plots, 
one has a similar asymptotical behavior near the asymptotes 
           \begin{figure}       
             \includegraphics[clip,width=7.0cm]{Fig5a_MRfq1-4ar08-04v4.eps}
          ~
\includegraphics[clip,width=6.0cm]{Fig5b-MqR10-12-14.eps} 

  \vspace{-0.4cm}
  \caption{{\small
        NS mass
        distributions $M(R)$, Eqs.~(\ref{MNStot}), (\ref{MNSV})
       and (\ref{MNSSfin}) 
          in Solar mass units,
          $M_{\odot}$, as
          functions of different variables.    
           (a): The  ES curvature radius $R$ dependence (in km) 
         is shown 
         by solid 
          lines for the
          relative mass densities,
          $\overline{\rho}/\rho^{}_{0}$=1~-~4,
         at the leptodermicity  $a/R=0.08$
          and incompressibility $\kappa=10$~.
          Dashed rare lines are the
         corresponding volume masses $M_{V}$, Eq.~(\ref{MNSV}),
         at $a/R=0$. 
         Frequent red and green dashed lines show 
         the masses $M(R)$ for the leptodermic
          parameter $a/R=0.04$
         and density values
         $\overline{\rho}/\rho^{}_0$=2 and 3, respectively.
         A sharp decrease of all solid (frequent dashed)
         lines is close to the asymptotes
                    near zero NS masses
                corresponding approximately to the $R_{\rm S}$
               values, Eq.~(\ref{RS}), for different
               radii $R$.
         Red (oblique cells) and orange (frequent points) spots show
         the observational data on
         the NS J0030+0441 from Refs.~\cite{TR19,MM19},  
         magenta (oblique lines) and cyan (vertical lines)
         spots correspond to
         the NS 
         J0740+6620
         \cite{TR21,AD24}. Blue (back oblique lines)
             and black (direct oblique cells) display the NS HESS J1731-347
             \cite{VD22}, 
                 and red (horizontal oblique lines) and
                 violet (direct oblique cells) show the data for the NS
                 Cen X-3 \cite{SN11,MR11,wiki:PSR-Cen-X-3},
         respectively; see
         Table \ref{table-1}.
         (b): The same as in (a) but as
         functions of
         the 
         inner mean  
         density
        $\overline{\rho}$ in units of the nuclear matter value
        $\rho^{}_0$ are shown
         for several values of the effective 
        radius 
        $R=10$ (black), $12$ (red), and $14$ (blue)
        km.
                  Dashed and solid
               curves are related
               to the values of
               the leptodermic parameters
               $a/R$=0.04 and 0.08.
Other parameters are the same as in (a).
}}
\label{fig5}
\end{figure}
 $Q_{\rm K}$ 
over the density variable $\overline{\rho}/\rho^{}_0$,
which correspond to $R_{\rm K}$ for the $R$ abscissa axis
in Fig.~\ref{fig3}.  The poles marked by arrows $Q_{\rm K}$, 
      modified by a rotation,
      are related mainly to 
      those of the  
      contributions shown in Fig.~\ref{fig2} by black
      solid curve; 
    cf.  with Fig.~\ref{fig3}.               
      A significant dependence
          of the MI on the relative crust thickness, $a/R$, is shown too (cf.
          solids with dashed and dotted lines).
           This dependence is
          the stronger the smaller
          NS radius $R$ of heavy NSs
              within the constraints mentioned above.  
 
              \begin{figure}
                \begin{center}
 \includegraphics[clip,width=10.0cm]{Fig6_PORRS_k2-10-50_finalissimo6.eps}
   \end{center}
 
   \vspace{-0.4cm}
  \caption{{\small
        NS periods $P$ (in units of ms) are plotted as functions
        of the dimensionless variable $R/R_{\rm S}$. 
            Straight line segments
        with different colors and boundaries 
        show different observational
        data for NSs presented in
        Table \ref{table-1} within the intervals of $R/R_{\rm S}$
        (Table \ref{table-3}; see
        its fifth column). 
             Dotted green line, and solid 
             and dotted
             black lines are
        the characteristic
         period $P_0$, Eq.~(\ref{omcond}), for 
        the full
        gravitational  metric with $\tau$, Eq.~(\ref{14tauinexpsol}),
        accounting for the $t,\varphi$ correlations  and surface
        contributions ``V+S''
        at the relative
        incompressibility values 
        $\kappa$=2 (weak, for a gravity 
        of the order of nuclear interaction), 
        10 (strong), 
        and 50 
         (super strong gravity),
        correspondingly;
        see Eq.~(\ref{kappa}) for $\kappa$. The cyan heavy dashed line shows
        the ``V+S'' Hogan $\tau$ calculations for $\kappa=10$.
                Dashed red line
        presents the 
        volume ``V''
        statistically averaged period $\tilde{P}_0^{(V)}$,
        Eq.~(\ref{P0Vt}). The relative crust thickness is $a/R=0.08$,
        and the effective mean radius is $R=12$ km. 
            Arrows show
        the corresponding asymptotes $R_{\rm K}/R_{\rm S}$.
   }}
\label{fig6}
\end{figure}

Figure \ref{fig5}(a) shows the mass distribution $M(R)$,
 Eqs.~(\ref{MNStot}), (\ref{MNSV})
        and (\ref{MNSSfin}), 
as function of the
effective radius ~$R$~
by using
only two  physical
parameters, the relative crust thickness $a/R$, 
and asymptotical inner
NS mass 
density,
~$\overline{\rho}/\rho^{}_{0}$= 1~-~4.
Using
a few parameters
fixed largely by observational
data,
in the leptodermic approximation,
    $a/R \ll 1$,
one can describe statistically averaged NS basic
properties of the
mass  $M$
as function of
the recently measured  
radius $R$; see
Refs.~\cite{TR19,MM19,TR21,AD24}.
 The surface effect is measured by the relative
difference
between the full NS mass, $M=M_{V}+M_{S}$ 
 (solid lines)
 and the
corresponding volume
component $M_{V}$ ($a=0$, dashed rare curves).
    A significance of these surface contributions
looks
similar to that in the Liquid Drop Model 
which describes 
the mean nuclear binding energies per particle 
depending on the
particle number. However, the main reason for
these contributions in NSs  is the interplay of the surface
    component with a strong gravitational field rather than the
    Coulomb interaction.
The NS mass $M(R)$ is not a monotonic
function of $R$ for 
the fixed inner
NS density,
$\overline{\rho}$, first of all
because of the surface component, $M_{S}$, Eq.~(\ref{MNSSfin}),
 modified
by the gravity.
This is
in contrast to the monotonic behavior of the volume mass,
$M_{V} \propto R^3$, in the Cartesian case of the small
Newtonian
gravitational
limit at $a=0$. 
For any given value of $\overline{\rho}$,
one finds a
rather pronounced maximum in dependence of the full mass $M(R)$,
in contrast to the total
 MI $\Theta(R)$ (Fig.~\ref{fig3})
but similarly as for statistically averaged MI component, 
$\tilde{\Theta}$ in Fig.~\ref{fig3}.
As well known \cite{RT87,LLv2},
the physical values of the NS radius $R$ for the 
Schwarzschild metric,
Eqs.~(\ref{smlim}) and (\ref{SchwarzIN}), have to obey
the equilibrium condition, $r_g < R < R_{\rm S}$,
where $r_g$, Eq.~(\ref{rg}), is the gravitational radius 
and $R_{\rm S}$ is the Schwarzschild
radius; see Eq.~(\ref{RS}).
Notice that $R_{\rm S}$ is a constant
independent of the
    radial coordinate $r$
because of the
leptodermic property assumed in the TOV derivations 
(see Refs.~\cite{RT39,RT87,OV39,MM24,MM25npa}.
The maxima are increased from about 1
to 2 $M_{\odot}$ for decreasing values of
$\overline{\rho}/\rho^{}_0$ from 4 to 1
at $a/R=0.08$,
respectively.  For $a/R=0.04$, one obtains larger 
maxima, about (2~-~3) $M_{\odot}$ for 
$\overline{\rho}/\rho^{}_0 = 3~-~2$.
  The NS mass for each of
these curves at a given 
value $\overline{\rho}$
disappears sharply 
in the limit $R \rightarrow R_{\rm S}$,
and does not exist at $R \geq R_{\rm S}$.
As mentioned above for the NS moments of inertia and 
 masses,
our derivations for the surface component of the NS mass, $M_{S}$,
 can not be immediately used near the point $R=R_{\rm S}$. 
As seen from Fig.~\ref{fig5}(a), our results are in
reasonable agreement with
the observational 
data, Refs.~\cite{TR19,MM19,TR21,VD22,AD24}.
For smaller NS masses,
$M/M_{\odot} \approx \{1.2~;~1.5\}$, for the NS
J0030+0451 \cite{TR19,MM19} with the slightly different radii,
$R=\{10.3~;~11.8\}$ km (orange spot) and
$R=\{11.6~;~13.1\}$ km (red spot),
one finds respectively good agreement
with our results 
for the 
inner mean density $\overline{\rho}=\{2~;~3\} \rho^{}_{0}$,
with $a/R$=0.08. A small mass of the NS HESS J1731-347,
$M/M_{\odot} \approx \{0.6~;~1.0\}$ with the
radii $R=\{9.6~;~11.3\}$ km (blue spot)
and  $R=\{10.9~;~11.8\}$ km (black spot), are associated also
with our results for
    $\overline{\rho}=\{2~;~3\} \rho^{}_{0}$,
       with the same $a/R$~, respectively.
     Notice that according to
    Ref.~\cite{VD22} and our calculations, the two results for
    the NS HESS J1731-347 in Table~\ref{table-1} are close, in
    spite of different model
    assumptions used in the
    NS spectra measurements. Two smaller close spots with the
        same radius
    $R=\{9.0~;~9.3)$ km and different masses, $M/M_{\odot}
    \approx \{1.0~;~1.4\}$ (red) and $\{1.4~;~1.6\}$ (violet spots)
    agreed well with green dashed lines assigned by 
    $\overline{\rho}=3\rho^{}_{0}$ and $a/R \approx \{0.04; 0.00\}$.
A large mass of the
NS 
J0740+6620 \cite{TR21,AD24},
$M/M_{\odot}=2.0~-~2.1$, and its radius $R=(11.3~-~13.6)$ km
(magenta and cyan spots),
 correspond to approximately the 
    same 
density,
$\overline{\rho}=(2~-~3) \rho_0$, 
but near a smaller 
    leptodermic parameter value $a/R$=0.04~.

Similarly as in Fig.~\ref{fig4}(b),
 Fig.~\ref{fig5}(b)
shows the NS masses $M$ as functions of the density $\overline{\rho}$
for the NS effective radii $R$=10~-~14 km, and two values of the leptodermic
parameters $a/R$. In line of the discussions of Fig.~\ref{fig4},
one obtains larger masses $M$ at smaller values of $a/R$
for all other fixed values of parameters. As expected, the masses $M$
are dramatically and non-monotonically changed below the
values of $\overline{\rho}/\rho^{}_0$
    related to the Schwarzschild radius
    $R_{\rm S}$, Eq.~(\ref{RS}), for any shown fixed parameters $R$ and $a/R$~.

\begin{figure}
  \begin{center}
  \includegraphics[clip,width=6.0cm]{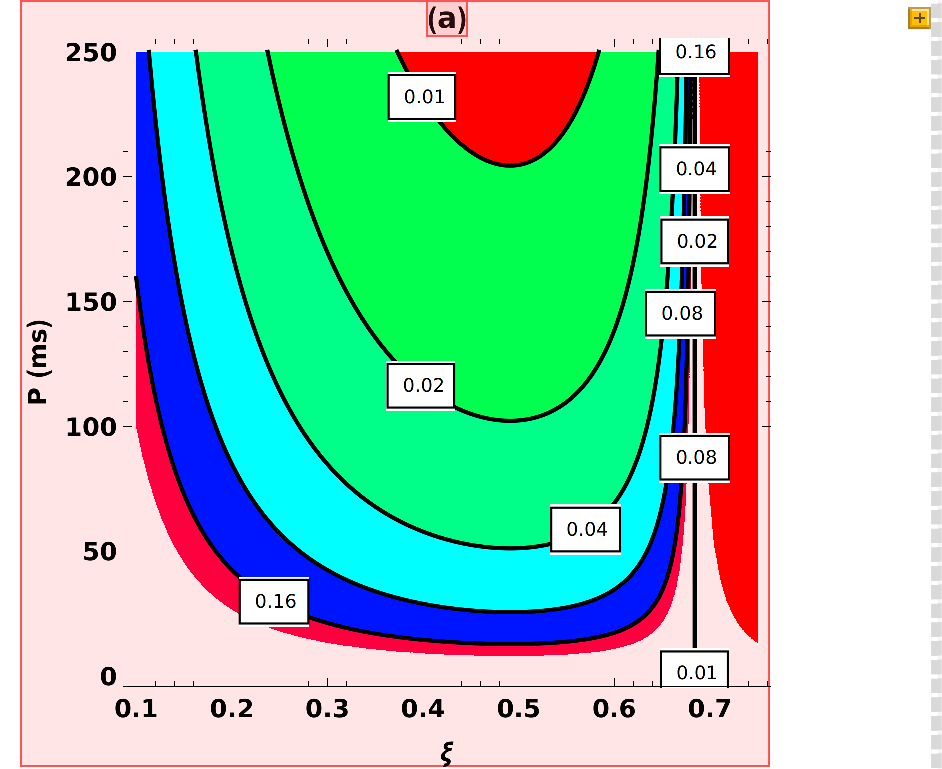}
  ~
 \includegraphics[clip,width=6.0cm]{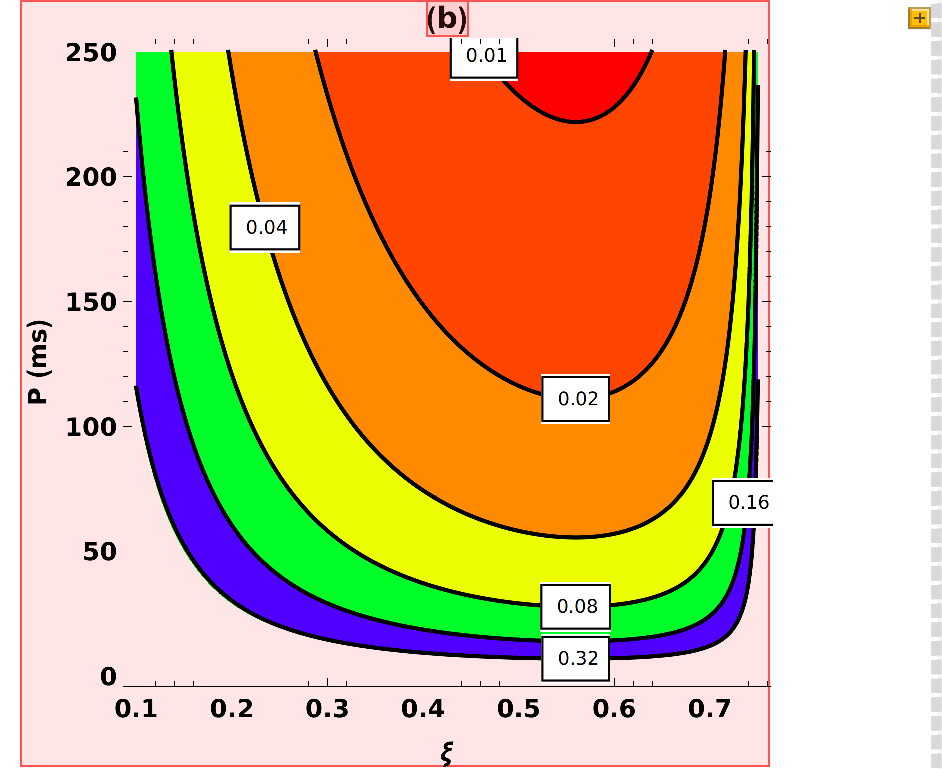}
  ~
  \includegraphics[clip,width=6.0cm]{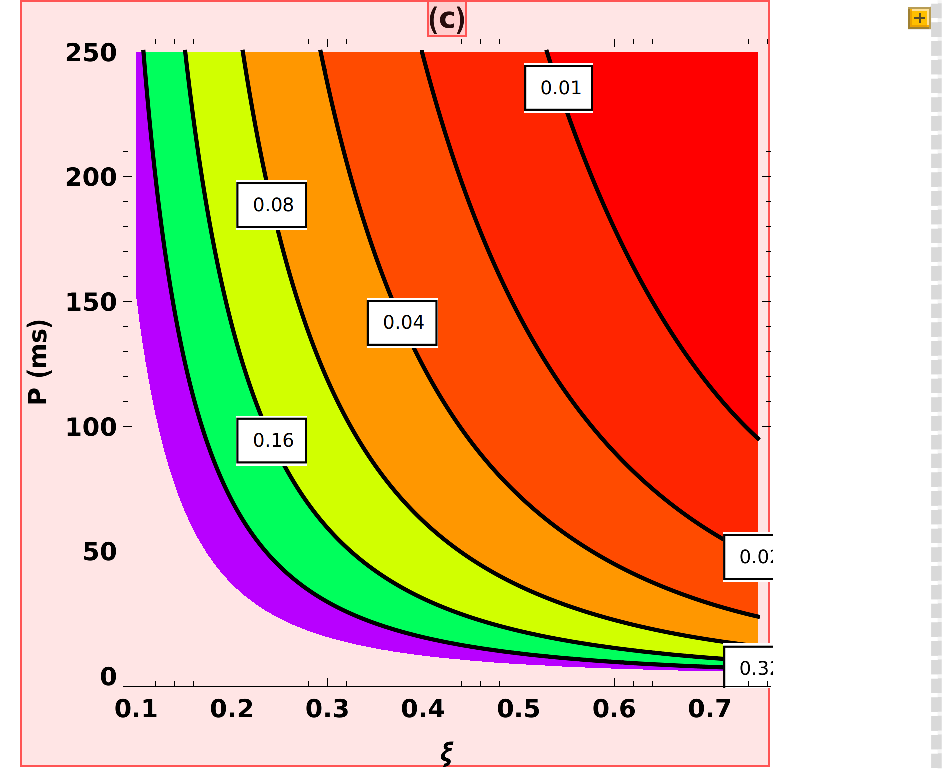}
  ~
  \includegraphics[clip,width=6.0cm]{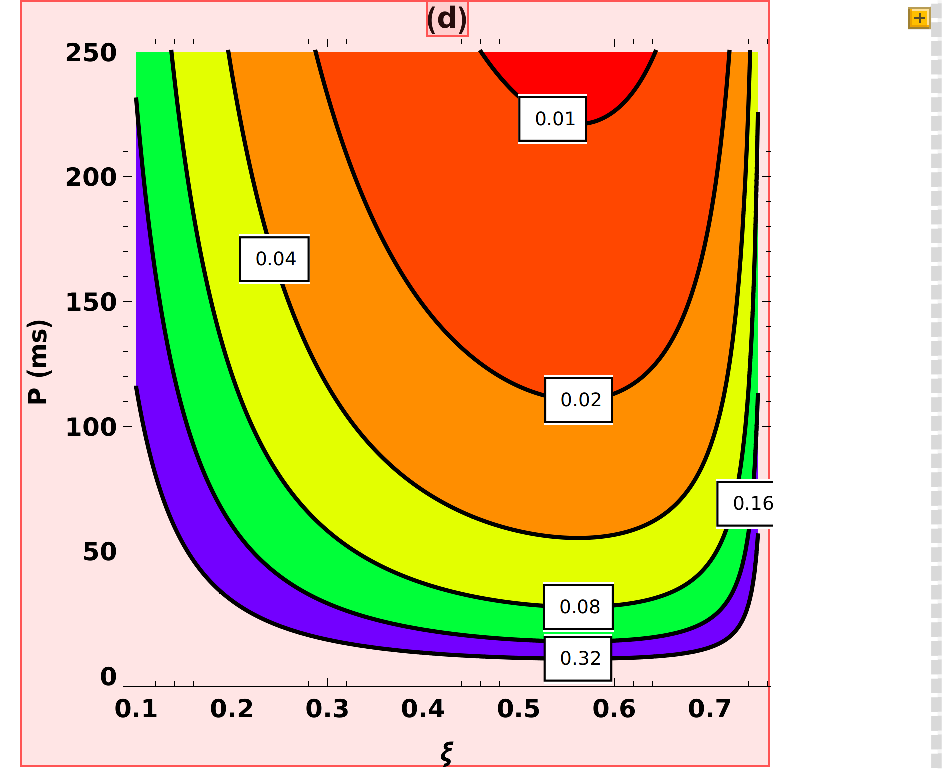}
  \end{center}

  \caption{{\small
      Contour plots show the 
      NS angular momentum parameter,
      $Ic/(M^2G)\approx 2\pi \Theta c/(M^2 P G)$, Eq.~(\ref{rotpar}),
      as function of the dimensionless variable
      $\xi=R/R_{\rm S}$ and period (in milliseconds) for the
      metric coefficient
      $\tau(r)$, Eq.~(\ref{14tauinexpsol}).
      The typical values of $Ic/(M^2G)$
      are shown in squares. (a): For incompressibility $\kappa=2$;
      (b): $\kappa=10$. 
      (c): $\kappa=50$.
      (d): Analytical Gauss approximation (a) in Table \ref{table-3} for
      $\kappa=10$.
          A mean radius is $R=12$ km, and a crust thickness $a/R=0.08$,
           as in Fig.~\ref{fig6}.
           Red areas on right of the vertical-line
           asymptote in (a) and
           white ones
           in (a,b,d) are related to non-physical
           parameters values.
          }}
\label{fig7}
\end{figure}
As shown in Table \ref{table-1}, 
      the radii $R$ of neutron stars above J0952-0607A
      are measured with a good accuracy
        simultaneously with masses $M$. 
        For other presented NSs we show the observed masses
            $M$ but the radii $R$
            were obtained by
        using the NS models.
      These  
      NS masses $M$
  are approximately in between 
  the values 
  0.6 and 2.5
 of the
 Solar
 mass $M_{\odot}$. 
  The lowest NS mass  
     is related to  ``a strange neutron star''
 HESS J1731-347
 \cite{VD22} while the largest one corresponds to the J0952–0607A.
   The NS radii $R$ in 
 Table~\ref{table-1}, measured simultaneously with NS masses $M$,
 are obtained within approximately a range
 $R \approx 7~-~13$ km (including errors). The smallest
     radius $R$ was obtained
     \cite{YK24} for
     another ``strange'' NS J1814, 
         while the largest radius was measured for
         the J0740+6620 NS.
         
     The
  observed rotation 
 periods $P$ (Table \ref{table-1}) are mainly
 much larger than
 the characteristic values of $P_0$, 
 calculated by Eq.~(\ref{omcond}),
 for a strong gravitational value of the
 incompressibility, $\kappa=10$, 
 Eq.~(\ref{kappa}).
    For  
   the discussed NSs with the observed radii $R$, they are 
   in good  agreement with the 
   adiabatic condition, except for NSs 4U 1608-52 and J0952-0607A;  see 
   Tables \ref{table-1}, \ref{table-2}, and
   Eqs.~(\ref{adcond}) and (\ref{omcond}).
   At the same time, the last column in Table \ref{table-2} 
 mainly agreed well
     with the results obtained for Hogan's 
     metric  (fourth column).
     We found even better adiabatic conditions for Hogan $\tau$ calculations.
     
    More accurate first-order perturbation condition, $Ic/(M^2G) \ll 1$,
      in our 
      calculations is well carried out for J1731-347, Cen X-3, 
      and Vela Pulsar, and partially, for Vela X-1,
       having larger rotation periods $P$, and even better
   for Hogan's case
   (see Table \ref{table-2}).
   Notice that $\overline{\omega}^2$ terms sometimes are
     significantly smaller 
     and can be  sometimes neglected, e.g.,
     for the NS J0437-4715.

Table \ref{table-2} shows the results using
the numerical integration
     for
     $\mathcal{I}_2$, Eq.~(\ref{IntI2}).  Table \ref{table-3} 
         presents different 
 analytical approximations
 in terms of $\mathcal{I}^{}_2$, Eqs.~(\ref{q2expJp}) for the Gauss 
 approximation (a);
 Eqs.~(\ref{q2exp})
 for its asymptote (b); and (\ref{q2as6exp2}) for
 the straight asymptotical solution (c)
 to the exact equation (\ref{14tauin}). As seen from
     Tables \ref{table-2} (third) and \ref{table-3} (second column),
     the Gauss analytical results
     (a) approximate mainly well our numerical calculations besides of the NS
     4U 1608-52, J0952-0607A, and partially, Vela X-1.
We found that within the shown digits, Appell and Gauss
     approximation,
     Eqs.~(\ref{apell}), is the same as given in (a).
     The approximation (b) (third) positive values of $Ic/(M^2G)$ are somewhat
     smaller 
     than those of (a) (second column) while there is essential
     discrepancy for negative ones of Table \ref{table-3}(a)
     and of the Table \ref{table-2} (third column).
     However, all values of (b) agreed well with those of (c), and both
     are extremely small.

The best 
 conditions for the linear perturbation
applicability, $ Ic/(M^2G)\ll 1$, among the shown NSs is 
those for the NS Cen X-3 with relatively a very large period
 (Tables \ref{table-1} and \ref{table-2}). The worst such
 conditions are found for
 the NS J0952-0607A, SAX J1748.9-2021, EXO J1745-268, and
 KS 1731-260, having the calculated radii $R$, all
 with relatively very small periods.
 We should especially notice here that for these NSs
 the rotational constraint
 $R<R_{\rm K}<R_{\rm S}$ fails ($R>R_{\rm K}$) at $\kappa=10$
 for mainly larger masses $M$ and smaller radii $R$.
 These conditions are violated
 also for the observed mass-radius measurements
 in the NS 4U 1608-52.
 According to Eq.~(\ref{MI}), 
     the MI $\Theta$, and hence,
 $Ic/(M^2G)$
     might be negative (non-physical), e.g., for
     larger mass $M$ and smaller radius $R$,
     see Table \ref{table-2} for a strong gravity $\kappa=10$.
     Notice also that for much larger
    gravity, e.g., a super gravitation $\kappa=50$, one finds
     positive MIs, instead of negative at $\kappa=10$, with
     essentially smaller values of $Ic/(M^2G)$ and $P_0$ for almost all NSs of
     this table.
     Taking the NS J0952-0607A as example,
     for this super gravity $\kappa$
     one obtains positive $Ic/(M^2G)$ values, but here
     $Ic/(M^2G)=\{0.59;~0.87\}$,
         i.e., the linear perturbation-approach condition still fails. For
     another interesting
     case, concerning a violation of the first-order perturbation approach
     NS J0740+6620 (version 1)
     at $\kappa=10$,
     one finds  at significantly larger value $\kappa=50$  much smaller
     values of
     $Ic/(M^2G)=\{0.32~;~0.60\}$, especially for quadratic terms.
                Generally speaking, the
     LPT condition, $ Ic/(M^2G)\ll 1$, is improved
     with increasing the incompressibility $\kappa$ from a strong
     ($\kappa=10$) to a super strong ($\kappa=50$) spherically symmetric
     case.
     
           We should emphasize that the results for $Ic/(M^2G)$
         in the case of the NS Vela X-1
         with extremely
         large period (Table \ref{table-1}) 
         (see Tables \ref{table-2} and \ref{table-3})
         are too small in absolute value, and we have to
         improve the calculation at $\kappa=10$ taking into account the
         NS surface deformation. However, in the spherical case ($\kappa=10$),
         our analytical (b) and (c) approximations for this NS
         (see Table \ref{table-3}) yield very good results,
         even much better
         than for the NS Cen X-3. All other $Ic/(M^2G)$ calculations
         in the super gravity NS Vela X-1 case for $\kappa=50$
         are much improved also, $Ic/(M^2G)=\{2.2;3.0\}\cdot 10^{-6}$
         for numerical 
         and $Ic/(M^2G)=\{1.1;2.3\}\cdot 10^{-6}$
         for analytical (c) 
         results for the third columns in Tables \ref{table-2} and
         \ref{table-3}, respectively.
         
         Figure~\ref{fig6} displays
      the characteristic limit periods  
      $P_0$,  Eq.~(\ref{omcond}),  
      as function of the ratio of the
      effective radius $R$ 
      to  the Schwarzschild radius 
      $R_{\rm S}$, Eq.~(\ref{RS}),  $R/R_{\rm S}$,
      versus the results calculated   
      for the data on the mass $M$, radius $R$
      and period $P$ for all NSs in
      Table \ref{table-1}. We present
      them at
      the definite periods $P$
         because of good accuracy of these 
         measurements, 
             besides of the NS EXO J1745-268 where we neglect
             small period errors for simplicity.
The intervals of the variable $R/R_{\rm S}$
         (fifth column in Table~\ref{table-3})
         are presented by line segments 
         within the 
             errors for the mass $M$ (second) and radii $R$ (third 
                 columns) in
         Table \ref{table-1}. 
         For the calculation of $P_0(R/R_{\rm S})$,
         this variable is expressed 
       through the mass $M$ 
       and radius $R$ by using both the  
       boundary outer-inner slitching condition,
       Eq.~(\ref{boundcond1}),
        and the gravitational radius 
           $r_g$ value, Eq.~(\ref{rg}),
           $R/R_{\rm S}=[2GM/(Rc^2)]^{1/2}$.
    The radius $R$ is taken 
        approximately
  as a mean 
    radius for NSs in Table \ref{table-1}, $R=12$ km. 
  The full characteristic rotational periods 
     $P_0$ (solid black) are compared also with 
    the volume statistically averaged $\tilde{P}^{(V)}_0$ 
    [dashed red line, Eq.~(\ref{P0Vt})]
    approximation.
    As the MIs $\Theta$, the limit periods $P_0$, Eq.~(\ref{omcond}),
    are dramatic 
       functions
      of the relative effective NS radius, 
      $R/R_{\rm S}$, because of the
      corresponding asymptotes 
      $R=R_{\rm K}$, 
      similarly as in Fig.~\ref{fig4}(a). 
      The critical periods  
      $P_0$ for $\kappa=2~-~10$ are
                    monotonically increasing functions of 
      $R/R_{\rm S}$, which converge to the corresponding 
      asymptotes $R=R_{\rm K}$. This is in
          contrast to mainly a decreasing behavior 
          of $P_0$ at $\kappa=50$ (dotted black line) up to
         $R/R_{\rm S}\approx 0.83$~.
    For the statistically averaged  dashed-red curve, one finds a maximum 
     for 
    the period $\tilde{P}_0^{(V)}$, with a cusp in the top 
     as 
     explained in section \ref{NSMI} (see Fig.~\ref{fig2}).
     In Fig.~\ref{fig6} we show also the essential dependence  of
      the full limit periods $P_0$
      on the dimensionless incompressibility parameter $\kappa$
      [Eq.~(\ref{kappa})]
  through the tension coefficient $\sigma$, 
 Eq.~(\ref{sigma1}),
 in the surface MI and energy contributions. 
 The values of $P_0$
 decrease with $\kappa$ from a 
 strong ($\kappa=10$) to super strong ($\kappa=50$)
 gravitation for the radius $R$ smaller than
     that for the black arrow
 $R_{\rm K}/R_{\rm S}$.
 As a result, for a strong $\kappa=10$ (solid black)
the adiabatic condition, Eq.~(\ref{omcond}), is carried out well
 for
 the most of neutron stars
 shown in Table \ref{table-1}, 
 except for those having the negative MI or  $Ic/(M^2G)$, e.g.,
 for the NS Vela X-1, J0952-0607A, and KS 1731-260. For stronger gravity
 ($\kappa=50$), this condition is fulfilled even better for all NSs,
 in contrast to the case of
 a weak gravity ($\kappa=2$).
       The surface effect is 
      enhanced because of the correlation contribution
      $\mathcal{T}_{t\varphi}$ 
      due to a change of the rotational root
      $R=R_{\rm K}$ [Eq.~(\ref{RKpole})], as seen from Figs.~\ref{fig6}
      and \ref{fig2}. Notice also 
          a quantitative agreement of these results
      for our calculations of $\tau$, Eq.~(\ref{14tauinexpsol}), with those
      for Hogan's metric (see Ref.~\cite{AM25npae}) for $R/R_{\rm S} \siml 0.7$,
      as well as for the approximations (b) and (c).
      The later agrees even a little better with the observation data
      for the same $\kappa=10$ (cf. with Fig.~\ref{fig2}).
      We found approximately in this region a reasonable agreement of
      the analytical
      Bessel solution for
      $\tau$, Eq.~(\ref{14tauinexpsol}), with the result of its exact
      calculations from Eq.~(\ref{14tauin}); see Fig.~\ref{fig1}.
      In this respect, the approximation (c) in Table~\ref{table-3} and
      Fig.~\ref{fig2}
      seems rather useful.   

       Thus, as clearly seen from Fig.~\ref{fig6} 
      and Tables \ref{table-1}, \ref{table-2}, and \ref{table-3}
      the adiabaticity condition (\ref{omcond}) 
      is carried out well
      for almost all NSs with observed masses and radii, even at 
        strong gravity  $\kappa=10$. 
          The comparison becomes 
       even better with increasing gravitation.
             In line of the results for the dimensionless
       angular momentum $Ic/(M^2G)$ (Tables \ref{table-2} and \ref{table-3}),
       it would be nice to improve our results taking into account 
       non-adiabatic effects. They are related, e.g.,
       to the quadratic frequency corrections
       $\sim \overline{\omega}^2$,  
           for several 
           NSs, 
           especially for
           the rotation of neutron stars 
               J0740+6620, 4U 1608-52, 4U 1820-30,
           and J1814 with observable masses and radii, and perhaps for
           almost all NSs with model-evaluated radii and extremely short
           periods.
                  In this non-adiabatic approach,
                  the NS MI $\Theta(\omega)$ depends essentially on the
                  rotation
                  frequency $\omega$. Another suggestion to improve
                  our results might be helpful by accounting for the NS surface
                  deformations.                 

            Contour plots in Fig.~\ref{fig7} 
       show the rotational
parameter, $\overline{\omega}(R/R_{\rm S},P)$, Eq. (\ref{rotpar}), depending on
the dimensionless variable $\xi=R/R_{\rm S}$ and rotational
period $P$ within the intervals which mostly
overlap data presented in Table \ref{table-1}. As seen
from the contour plots (a) and (b), for the 
    incompressibility $\kappa$ up to
$\kappa=10$, the parameter $\overline{\omega}$ has a
minimum.
With further increasing $\kappa$, at a super strong gravity
$\kappa=50$ for any given period value $P$,
one ﬁnds a monotonic behavior up to $\xi\approx 0.75$,
i.e., the minimum
disappears in Fig.~\ref{fig7} (c). 
    The analytical approximation (a) (Table
\ref{table-3}) in (d) is almost identical to the numeric calculations (b).
 The rotation parameter $\overline{\omega}$,
Eq.~(\ref{rotpar}), in (a,b,d) has a minimum because 
of the maxima of the NS mass $M$,
in spite of a monotonical behavior of the MI
$\Theta$ within the allowed physical values smaller than
the 
rotational root $R_{\rm K}$ [Eq.~(\ref{RKpole})]. With growing the eﬀective
radius $R$ for the same ﬁxed period $P$ and incompressibility
$\kappa=50$ (super gravity), one obtains basically 
smaller values
of $\overline{\omega}$. As seen from all these contour plots, the values of 
$\overline{\omega}$ are mainly small,
    except for the areas of very small
periods and very large $\xi$. 

  In line of all these discussions
     we may conclude that the
     linear perturbation approximation fails in the case of
     $\overline{\omega} \simg 1$ (or formally, negative
     $\overline{\omega}$)
         for several neutron stars with their very short periods.
     Therefore, in  this case we
     should study
     a non-linear (non-adiabatic) approach, taking into account, e.g., the
     centrifugal $\overline{\omega}^2$ corrections, or NS surface
     deformation, that might be a
     challenge for
     a forthcoming work.
 
     \section{Conclusions}
\l{concl}
     
The macroscopic effective-surface approximation based on the
leptodermic expansion
over a small parameter - a ratio of the 
crust thickness $a$
     to the effective radius $R$ of the system - is extended for
     the description
      of basic mean neutron-star (NS)
      properties for small rotational frequencies $\omega$.
      Starting from the Kerr metric approach in the 
       outer Boyer-Lindquist 
      and inner Hogan 
      coordinates, in the linear perturbation
      approximation over the dimensionless angular momentum parameter,
      $I c/(M^2 G)$  we
      solved analytically
          the 
          GRT
          equation for the off-diagonal $g^{}_{t\varphi}$ component
          of the gravitational metric. 
              The two-branch outer-inner
          solution
          for the rotational metric is continuous through the
          effective NS surface 
              at the statistical equilibrium, in accordance with that
          for the non-rotating
          Schwarzschild metric.
 Our NS macroscopic equation of state, 
 $\epsi=\epsi(\rho)$, is formulated in a simple but 
 general form  as a sum of the volume and 
 surface (gradient) terms for different inter-particle interactions, 
 modified by a strong 
 gravitational field.
  The gravitational part of the energy density, 
  $\mathcal{U}(\rho)$, in our approach inside the NS
  is taken into account 
  in the simplest quadratic form of the expansion
  of $\mathcal{U}(\rho)$ over powers of
 the density differences $\rho-\overline{\rho} $ , where
        $\overline{\rho}$ is the inner
  mean NS
  density.  This density  is assumed to be
  in a few times larger than, 
      or of the order of the mass 
 nuclear-matter density.
   Our results for the dependence of the NS mass $M$ on
          the NS radius $R$, including  
          essentially
          the gravitational and surface effects
          are in a
          reasonable agreement with their
          recent
          observational  data for
         several neutron stars, for which the masses and radii are
         recently measured simultaneously.
  The moments of inertia $\Theta$
 for NS as a rotating perfect dense-fluid system are derived
 in terms of the statistically averaged MI, $\tilde{\Theta}$,
 and time-angle $t,\varphi$ correlation contribution, $\mathcal{T}_{t\varphi}$,
 which are, in turn,  the sums of the volume and surface terms.
 Taking into account the approximate self-consistency relation
 between the NS
 angular momentum  $I$ and 
 the $t,\varphi$ gravitational metric component $g^{}_{t\varphi}$ and a feedback
 to the angular momentum $I$, one obtains a non-linear expression for the
 MI 
having a pole.
 Therefore, due to this  rotational gravitational coupling, one finds 
  additional constraints over NS radii $R$, which are due to the rotational
      perturbation, $ R < R_{\rm K} < R_{\rm S}$, where
      $R_{\rm K}$ is the root of the MI denominator, 
      $\mathcal{T}_{t\varphi}(R)-1$.
      The root $R_{\rm K}$ depends essentially on the surface component
      of the correlation contribution $\mathcal{T}_{t\varphi}(R)$.
           The surface components of $\tilde{\Theta}$ and
      $\mathcal{T}_{t\varphi}$ are 
      expressed in terms of the tension coefficient $\sigma$ 
      of the NS surface 
      energy $E_S$. 
      We found 
  that 
  most of neutron stars 
      with observable mass $M$ and radius $R$
      and rotating with measured periods which are larger or of the order
      of about 5 ms 
  obey well the simple 
  macroscopic adiabatic conditions
      for a strong gravity.
        However, we found a dramatic 
        dependence of the MI 
        because mainly of the 
      $t,\varphi$ correlations coupling. 
         The simple adiabatic-rotation approach
      should  be essentially improved near the critical 
      asymptotical values of the effective 
      radius $R$  near the rotational boundary value $R_{\rm K}$,
              and for some NSs with extremely small periods of
      the order of millisecond.
      We found also the essential dependence of the MI 
      on other parameters, such as the surface tension 
    coefficient $\sigma$,
           through the leptodermic, $a/R$, 
          and  the nuclear-gravitational incompressibility,
      $\kappa$, 
      parameters.

    As perspectives, one can generalize  
   our analytical approach to take into account the NS surface deformations 
   and the NS quadratic rotational frequency corrections 
    to the moments of inertia. We are planning to take into account the
    fluctuations, and critical phenomena such as superfluidity,
    and to compare our macroscopic
    approach with 
    relativistic mean-field approaches in more details.
    It would be nice also to study
    the NS $\beta$ equilibrium,
    and its inner structure accounting for hadrons, and check the accuracy
    by accounting for a more general expression for the surface term of
    the energy density with an isotopic asymmetry. It is especially
    interesting to derive
   the rotational corrections to the
     TOV equations and compare them with the standard TOV approach.
   In particular, solving these
  equations for the pressure, one can take into account 
  the
  ES deformation due to the NS rotations 
 for a strong gravity. 
    
\ack

The authors greatly acknowledge 
V.Z.\ Goldberg,
M.I.\ Gorenstein, F.A.\ Ivanyuk,   
J.\ Holt, C.M.\ Ko, E.I.\ Koshchiy, J.B.\ Natowitz,
S.A.\ Omelchenko,
G.V.\ Rogachev, A.I.\ Sanzhur, Y.V.\ Shtanov, B.I.\ Sidorenko,
Yu.V.\ Taistra, 
V.I.\ Zhdanov
for many creative and useful discussions.

\begin{appendices}

\setcounter{equation}{0}
\renewcommand{\theequation}{A\arabic{equation}}

\section{Static ES approach}
\l{appA}

\subsection{
  Volume and surface (crust) density}
\l{appA1}

In the non-rotating leptodermic system volume,
asymptotically far away from the ES,
the terms
of Eq.~(\ref{eq}) containing 
derivatives of the
mass density
$\rho$ are assumed to be relatively small.
Therefore, 
neglecting gradient terms in Eq.~(\ref{eq})  
 in the system volume,
  and solving the simplified equation approximately
      up to quadratic terms over $\rho-\overline{\rho}$,
one obtains
\be\l{solvol}
\rho=\overline{\rho}\left(1 + \frac{9\mathcal{M}}{K_G}\right), 
\ee
where $\mathcal{M}$ is a small surface (capillary) correction
to the leading component
of the
chemical potential $\mu$ ($\mathcal{M}\propto a/R$;
    see Ref.~\cite{MM24}).

We will now   
solve
    the typical catastrophe
equation (\ref{eq}) for
the distributions of the 
density $\rho$ through the ES at leading order over a small 
parameter $a/R$. 
Multiplying Eq.~(\ref{eq}) at leading order by the 
derivative 
$\partial \rho/\partial r$, and integrating over $r$ with
the boundary conditions,
$\rho \to 0$ and $\rho^\prime \to 0$
for $r \to \infty$,   and using Eqs.~(\ref{vol})
and (\ref{surf}),
one obtains 
\be\l{eq0}
\frac{\d \rho}{\d r}=-\sqrt{\frac{ \eps^{}_G(\rho)}{\mathcal{C}}}~,
\ee
where 
$\eps^{}_G(\rho)$ is given by 
Eq.~(\ref{epsKG}).
Introducing for convenience
the dimensionless quantities,
\be\l{units}
y=\frac{\rho}{\overline{\rho}}~, \quad x=\frac{r-R}{a}~,
\quad\mbox{with}\quad
a=\sqrt{\frac{18m\mathcal{C}\overline{\rho}}{ K_G}}~,
\ee
    ($x$ in this Appendix should not be mixed up with different
    notations for $x$ in
the main text and Appendix \ref{appB}) one 
has from Eq.~(\ref{epsKG})
\be\l{varepsilon}
\eps^{}_G(\rho)\equiv
\frac{\overline{\rho}K_G}{18m}~\epsilon(y),~~\epsilon(y)=y(1-y)^2.
\ee
 Using these definitions, 
one can reduce Eq.~(\ref{eq0}) to  the 
following dimensionless form:
\be\l{eq0yx}
\frac{\d y}{\d x} =-\sqrt{\epsilon(y)}~.
\ee
From the asymptotes of the
explicit analytical expressions for the particle density
$y(x)$ at large $x$, one may see 
a typical behavior,
$y(x) \propto e^{-x}$~.
This is a clear reason in order to call $a$,
Eq.~(\ref{units}), the NS
 crust thickness parameter.

 A single boundary condition 
 for the unique solution of 
the first-order differential equation can be  derived 
within the leading
ES approximation from the definition
of the ES as the points of maximal values of the derivative, 
$\partial \rho/\partial r$ at $r=R$, namely, $y''(0)=0$ at $x=0$. 
Differentiating Eq.~(\ref{eq0yx}) over $x$,
for the position of
the ES, $y=y_0$ at $x=0$, one
obtains [see, e.g., Eq.~(\ref{varepsilon})],
\be\l{boundcond}
-\beta y^2\epsilon(y_0) +
y_0(y_0+\beta y_0^2)\frac{\d \epsilon(y_0)}{\d y}=0~. 
\ee
The function $\epsilon(y)$ is
given by the specific choice of the
 inter-particle NS interaction which leads to the energy density 
 Eq.~(\ref{enerden}), e.g.,
 associated with 
 the nuclear Skyrme and dense molecular van der
Waals interactions,
including the gravitational field.
The integration constant is determined by the boundary 
condition (\ref{boundcond}).
The leading-order solution of Eq.~(\ref{eq0yx})
for arbitrary $\epsilon(y)$
can be found
explicitly  
in the inverse form:
\be\l{sol0xy}
x=-\int_{y_0}^{y}\frac{\d w}{\epsilon^{1/2}(w)}~.
\ee
Using also Eq.~(\ref{varepsilon}) for $\epsilon(y)$,
this integral can be calculated analytically in terms of the elementary
functions 
within the condition
$a/R \ll 1$ at leading order.
For the example of simple solutions $x=x(y)$, Eq.~(\ref{sol0xy}),
of Eq.~(\ref{eq0yx}) with the boundary condition, Eq.~(\ref{boundcond}),
one finally obtains \cite{strmagden,MS09,MM24}
\be\l{b0g0}
y(x)=\tanh^2[(x-x_0)/2],\qquad
x<x_0=2 {\rm arctanh}(1/\sqrt{3})~,
\ee
and zero for $x\geq x_0$; see Ref.~\cite{MM24}.
The plots with the density proﬁle of the NS
with the Skyrme interaction KDE0v1 was obtained
by solving TOV equations numerically with some parameters
in agreement with the analytical
solution, Eq.~(\ref{b0g0}).

\subsection{Mean NS energies}
\l{appA2}

The total ETF energy $E$,
 Eq.~(\ref{energytot}), can be approximated as (Ref.~\cite{MM24})
\be\l{enertot}
E \approx E_V+E_S~,
\ee
where
$E_V$ is the volume part of the total energy,
\be\l{enertotv}
E^{}_V=\overline{\rho} c^2\int \d \mathcal{V}
= 2 \pi \overline{\rho} c^2
R_{\rm S}^3 f\left(R/R_{\rm S}\right).
\ee
For the
surface part, one obtains
\be\l{enertots}
E_{S} \approx \sigma S~, 
\ee
 where $S=4\pi R^2$ is the surface area value, and $\sigma$
 is the leading-order tension coefficient, 
\be\l{sigma1}
\sigma=\frac{a \overline{\rho} K_G}{9 m} \mathcal{J}(R)
\int_{0}^1 \d y
\sqrt{\epsilon(y)} \approx
\frac{4a \overline{\rho} K_G}{135m} J(R)
=\frac{16}{45}~a \kappa \overline{\rho}c^2 \mathcal{J}(R);
\ee
see Eq.~(\ref{Jin}) for $\mathcal{J}(r)$ at $r=R$.
Within the quadratic approximation for $\epsilon(y)$, 
Eq.~(\ref{varepsilon}), one finally obtains the
two right explicit expressions for Eq.~(\ref{sigma1}).
For nuclear physics ($\mathcal{J}(R)=1$), the tension coefficient $\sigma$
is related to the   Skyrme interaction constants, in good agreement
    with the van der Waals capillary theory \cite{RW82}.
    Notice that we neglected  the contribution $\sigma_{\rm corr}$
    to the tension coefficient $\sigma$,  
    \be\l{sigmacorr}
    \sigma_{\rm corr}=\frac12 a\overline{\rho}c^2 \mathcal{J}(R)~,\qquad
     \frac{\sigma_{\rm corr}}{\sigma}
      \approx \frac{135}{96~\kappa}~.
    \ee
    This is relatively small correction to the main tension
    coefficient (\ref{sigma1}) under the adiabatic
    condition for a strong gravitational incompressibility,
    $\kappa \ll 1$,
    Eq.~(\ref{kappa}).

\setcounter{equation}{0}
\renewcommand{\theequation}{B\arabic{equation}}

\section{Gravitational metric solutions for NS rotations
      as a linear response}
\label{appB}

For calculations of the correlation
contribution $\MItf_{t\varphi}$, Eq.~(\ref{MItf}), one has to calculate
the factor $\tau(r)$ within the LPT approach.
For the angular momentum $I$ in the case of the stationary rotation
of the perfect dense-fluid system around its symmetry axis,
one has 
Eq.~(\ref{Igen}); see 
Refs.~\cite{LLv2,LLv6,MV08,BL67,PH76,FIP86,NS03,AW08,NS13,NS17}.
The
gravitational metric $g_{\mu\nu}$ for a small rotational frequency 
parameter, $\Omega \propto \overline{\omega}$
[Eq.~(\ref{rotpar})], 
up to $\overline{\omega}^2$ terms, can be
presented
as
\cite{MV08,BL67,PH76,AW08}
\be\l{AWmet}
{\rm d} s^2=e^\nu c^2{\rm d} t^2
+2\tau \Omega \sin^2\theta c\d t \d \varphi
-
e^\lambda{\rm d} r^2-
     r^2 {\rm d}\theta^2-
     r^2 \sin^2\theta {\rm d} \varphi^2~,
     \ee
where $\nu(r)$, and $\lambda(r)$ are 
the Schwarzschild parameters \cite{RT87}.
To derive the equation for $\tau$
from the GRT
    equation  (at zero cosmology constant),
    linear over the scalar curvature tensor,
one should first calculate the Christoffel
symbols and Ricci tensor \cite{LLv2}.
Using the LPT, for the graviational metric $g$, one can
write
$g=g^{}_0+g^{}_1~$,
where $g^{}_0$ is the two-branch outer-inner Schwarzschild solution
\cite{RT87},
Eqs.~(\ref{smlim})
and (\ref{SchwarzIN}), and
$g^{}_1$ is the first order rotational correction over $\overline{\omega}$.
Within this LPT for an axially symmetric system,
one can look for the solutions
$g_{\mu\nu}$ of the GRT equation
as
\be\l{gmnW}
\!\!\!g^{}_{00}\!=\!e^\nu,\quad g^{}_{11}\!=\!-e^\lambda,\quad
g^{}_{22}\!=\!-r_{}^2,
\!\!g^{}_{33}\!=\!-r_{}^2\sin_{}^2\theta~,\quad
g^{}_{03}\!=\!g^{}_{30}\!=\! 2 \tau \Omega \sin^2\theta~, 
\ee
\noindent where $\nu=\nu(r)$, $\lambda=\lambda(r)$ 
    are known as the Schwarzschild
    gravitational metric parameters [see, e.g.,
    Refs.~\cite{RT87}  at zero order
($\overline{\omega}=0$)], and $\tau=\tau(r)$
is unknown function of the radial coordinate $r$ in the spherical
coordinate system $t,r,\theta,\varphi$;
$\Omega \propto \overline{\omega} $ 
is 
Kerr rotational parameter, Eq.~(\ref{OmegaK}),
and  $g^{}_{\mu\nu}$ for all other $\mu$ and $\nu$
subscripts are zeros.
Calculating now the Christoffel symbols
with the help of
Eq.~(\ref{gmnW}), 
one obtains the Ricci tensor, $\mathcal{R}_{\mu\nu}$, in the same linear
approximation over 
$\overline{\omega} \propto \Omega $, 
\bea\l{RmnW}
\mathcal{R}_{00}&=&\frac{1}{4r}e^{\nu-\lambda}~
\left[2r \nu^{\prime\prime} +\left(4-r\lambda^\prime\right)\nu^\prime
  +r\left(\nu^\prime\right)^2\right],\qquad
\mathcal{R}_{11}=-\frac14\left[\left(2\nu^{\prime\prime}+
\left(\nu^\prime\right)^2\right)+\frac{1}{r}\lambda^\prime
\left(4+r\nu^\prime\right)\right],\nonumber\\
\mathcal{R}_{22}&=&\frac12 e^{-\lambda}
\left[r\left(\lambda^\prime-\nu^\prime\right)+2e^\lambda-2\right]~,\qquad
\mathcal{R}_{33}= \sin^2\theta~\mathcal{R}_{22}~ ,\nonumber\\
\mathcal{R}_{03}&=& \frac{\Omega}{2r^2} \sin^2\theta~e^{-\lambda}
\left\{r^2\left[2\tau^{\prime\prime}-
  \left(\lambda^\prime + \nu^\prime\right)\right]\tau^\prime 
\quad\!\!+4\left(r \nu^\prime -e^\lambda\right)\tau\right\}~.
\eea
All others metric elements $\mathcal{R}_{\mu\nu}$ are zero. 
Notice that two off-diagonal 
(equivalent) metric elements,
$\mathcal{R}_{03}=\mathcal{R}_{30}\propto \Omega$, appear because
of the rotational off-diagonal  elements of the gravitational metric, 
$g^{}_{03}=g^{}_{30}\propto \Omega$.
Calculating also the trace invariant,
 one obtains the same three equations
 for the diagonal 
     (``00'', ''11'', and ``22'' which is equivalent to ``33''
 due to the spherical symmetry) components independent of $\Omega$
  at zero order. 
 However, one more equation for the off-diagonal
  ``03'' component 
 (proportional to $\Omega$) 
 appears,
 \be\l{14tau}
 r^2\tau^{\prime\prime} - \frac{r^2}{2}\left(\nu^\prime+\lambda^\prime\right)
 \tau^\prime -\tau\left[r^2\nu^{\prime\prime}
   -r\lambda^{\prime}\left(\frac{r}{2}\nu^\prime+2\right)
+\frac{r^2}{2}\left(\nu^\prime\right)^2+2\right]=0~.
 \ee
 \subsection{Outer solutions}
 For the outer case, $r>R$, according to Eq.~(\ref{nulamout}), 
     one has
 $\nu^\prime+\lambda^\prime=0$. Therefore,
 Eq.~(\ref{14tau}) is simplified to
 \be\l{14taur.1}
 r^2\tau^{\prime\prime} 
 - 2 \tau=0~.
 \ee
 Introducing the dimensionless variables,
 \be\l{xyvar}
  x=r/r_g,\quad y(x)=r_g^2 \tau(r)~,
  \ee
  Eq.~(\ref{rg}),
  one can  
 re-write Eq.~(\ref{14taur.1}) as
  \be\l{14taux.1}
  x^2y^{\prime\prime}(x)-2y(x)=0~.
  \ee
 A general
  solution of this equation, obtained for $r>R$ 
   is given by
  Eq.~(\ref{14tauxsolgen}).

   \subsection{Inner solutions}

   Substituting Eq.~(\ref{nulamin}) into Eq.~(\ref{14tau}), one obtains
   the equation for $y(x)=R_{\rm S}^2\tau(x)$,
   $x=r/R_{\rm S}$ in
   the inner region, $r<R$,
   \be\l{14tauin}
 a^{}_2y^{\prime\prime}(x)+a^{}_1y^{\prime}(x)+a^{}_0 y(x)=0~,
   \ee
   where
\bea\l{acoef}
   &\hspace{-0.7cm}a^{}_0 =
2\left[6 \sqrt{1-\xi^2}-(10-9\xi^2)\sqrt{1-x^2}
  \!\!+
  x^4\left(21\sqrt{1-\xi^2}-4\sqrt{1-x^2}\right)\right.
     \nonumber\\
     &\!\!-\left.x^2\left(27\sqrt{1-\xi^2} 
     -(32-27\xi^2)\sqrt{1-x^2}\right)\right],\nonumber\\
   &\hspace{-3.0cm}a^{}_1 =3x^3\left[\sqrt{1-\xi^2}\left(1-x^2\right)
  -3\left(1-\xi^2\right)\sqrt{1-x^2}\right],\nonumber\\
   &\hspace{-0.7cm}a^{}_2 =-x^2(1-x^2)\left[6\sqrt{1-\xi^2}-
     \left(10-9\xi^2\right) 
     \sqrt{1-x^2} -
      x^2\left(6\sqrt{1-\xi^2}
     -\sqrt{1-x^2}\right)\right],
     \eea
   and $\xi=R/R_{\rm S}$.
   This equation can be solved for $y(x)$ in terms of the analytical
   asymptotic
   expression as function of a small $x$ up to 6th order at any
   $0<\xi=R/R_{\rm S}<1$ with two integration constants. Keeping
   only non-singular solution going to
   0 at $x\rightarrow 0$, one obtains
   \bea\l{as6}
   &&\!\!\!\!y\approx c^{}_1x^2\left\{\!-
   \frac{6 \sqrt{1 - \xi^2} x^2}{5 (9 \xi^2-8)}
     + \frac{x^4}{
     35 (-8 + 9 \xi^2)^2} 
      + \frac{3 \sqrt{1 - \xi^2} x^4}{35 (9 \xi^2-8)^2}
     - \frac{
       3 \sqrt{1 - \xi^2} x^4}{35 (9 \xi^2-8)}\right.\nonumber\\
  &&  \!\!\!\!\!\left. -\frac{1}{70} \!\left(\!14 x^2
     \!+\! 5\!x^4-\!70\right)
     \!-\!
     \frac{28 x^2 \!+\! 3 x^4}{70 (9 \xi^2\!-\!8)}\right\}. 
     \eea
            To simplify Eq.~(\ref{14tauin}) for analytical solutions
          we expand
   the coefficients, Eq.~(\ref{acoef}), in power series, first in small
   $\xi$, and then, therefore ($r\leq R$, or $x\leq \xi$) in small
   $x=r/R_{\rm S}$ , both up to quadratic terms. Finally, one obtains 
   \be\l{acoefexp}
   a^{}_0 =-8 + 20x^2+12 \xi^2,\quad
       a^{}_1 =0~,\quad a^{}_2 =4x^2~.
   \ee
   With these coefficients,
   one arrives at
   much simpler equation:
   \be\l{14tauinexp}
   x^2y^{\prime\prime}(x) -\left(2-5x^2-3 \xi^2\right) y(x)=0~.
\ee
This equation has the following solutions
: 
\be\l{14tauinexpsolgen}
y(x)=\sqrt{x}\left[J_{p}\left(\sqrt{5}x\right)c^{}_1+
  Y_{p}\left(\sqrt{5}x\right)c^{}_2\right]~,
\ee
where $J_{p}(z)$ and $Y_{p}(z)$ are the Bessel functions of the order
$p$, Eq.~(\ref{p}),  with the argument $z=\sqrt{5}x$, 
$c{}_1$ and $c{}_2$ are arbitrary integration constants. The coefficient
$c{}_2$ should be put zero
because of the finiteness condition for $\tau$ in the limit $x \rightarrow 0$
($r \rightarrow 0$).
Therefore, the solution for $\tau$ is given by Eq.~(\ref{14tauinexpsol}).

Expanding the Bessel function $J_p(\sqrt{5}x)$ in  the integrand,
Eq.~(\ref{IntI2}), one can analytically integrate over $x$.
For $\mathcal{I}^{}_2$,
one obtains
\bea\l{apell}
&&\mathcal{I}^{}_2=\frac{2^{-2-p}~5^{p/2}x^{p+7/2}}{2 A^2~(1-4 A^2)^2~(7+2p)
  \Gamma(2+p)}
\left[(1-4A^2)~(1+20 A^2-4 p)\right.\nonumber\\
  &&\left.\times \mbox{AppellF1}\left(\frac{7}{4}+\frac{p}{2},
  -\frac{1}{2},1;\frac{11}{4}
+\frac{p}{2}; x^2;\frac{x^2}{1-4A^2}\right)\right.\nonumber\\
&&+ \left.8A^2~(20 A^2+4p-1) 
\mbox{AppellF1}\!\left(\frac{7}{4}+\frac{p}{2},
-\frac{1}{2},2;\frac{11}{4}
+\frac{p}{2};x^2;\frac{x^2}{1-4A^2}\right)\right.\nonumber\\
&&+\left.(1-4A^2)^2~(4p-1)
F\left(\frac{7}{4}+\frac{p}{2},\frac{1}{2};
\frac{11}{4}+\frac{p}{2};x^2\right)
+80 A^3 (1-4A^2)
F\left(\frac{7}{4}+\frac{p}{2},1;\frac{11}{4}+\frac{p}{2};
\frac{x^2}{1-4A^2}\right) \right.\nonumber\\
&&+\left.
+16A^3(20 A^2+4p-1) 
F\left(\!\frac{7}{4}+\frac{p}{2},2;\frac{11}{4}+\frac{p}{2};
\frac{x^2}{1-4A^2}\right)\right],
\eea
where $\mbox{AppellF1}\!\left(\alpha,\beta,\gamma;\delta;z;w\right)$ is the
Appell (double hypergeometric series)
function,
$F\left(\alpha,\beta;\gamma;w\right)$ is the Gauss (one hypergeometric series),
$p$
and $A$ are functions of $\xi=R/R_{\rm S}$,
Eqs.~(\ref{p}) and (\ref{Tolconsschwm}), respectively.
Independently, using the 
expansion of the Bessel function
$J_p(\sqrt{5}x)$ and a more smooth quantity $\sqrt{1-x^2}$
up to the same second order
over $x$ [used already also in the derivation of Eq.~(\ref{IntI2})]
in  the integrand of
Eq.~(\ref{IntI2}), one can obtain  a more simple expression for
$\mathcal{I}^{}_2$
in terms of only three Gauss functions $F(n,\alpha,\beta,w)$, where
$n=1,2$:
\bea\l{q2expJp}
&&\mathcal{I}_2=\frac{2^{-p} 5^{p/2} \xi^{7/2 + p}}{
  (2A-1)^2 (4 A^2 (1 + p) (7 + 2 p) \Gamma(1 + p)} 
\left[(2 p-3)
      F\left(1, \frac{7}{4} + \frac{p}{2}; \frac{11}{4} + \frac{p}{2};
      \frac{\xi^2}{2 (1 - 2A)}\right)\right.\nonumber\\
&&+ \left.(2 p-3)(2A-1)^2 F\left(1, \frac{7}{4} + \frac{p}{2}; 
      \frac{11}{4} + \frac{p}{2}; \frac{\xi^2}{2}\right) \right.\nonumber\\
&&      \left.
    +  2A (10 A + 2 p-3)
      F\left(2, \frac{7}{4} + \frac{p}{2}; \frac{11}{4} +
       \frac{p}{2}; 
       \frac{\xi^2}{2 (1 - 4 A)}\right)\right],
   \eea
   with the same parameters $p$ and $A$ [Eqs.~(\ref{p})
   and (\ref{Tolconsschwm})].

  \setcounter{equation}{0}
\renewcommand{\theequation}{C\arabic{equation}}

\section{Derivations of macroscopic NS moments of inertia within
  Hogan's approach}
\label{appC}

For the 
volume $t,\varphi$ correlation
contribution $\mathcal{T}_V$, Eq.~(\ref{MItf}), with Eq.~(\ref{tauin})
for Hogan's $\tau$, $\mathcal{T}_V^{(H)}$,  one obtains
[see Eq.~(\ref{adMItfV})]
\bea\l{I2tVH}
\mathcal{T}^{(H)}_V \approx
W_2^{(H)}(z^{}_0)&=& \frac{2}{\left(1-z^{2}_0\right)^{3/2}}
\int_{z^{}_0}^1 \d z
\frac{(1-z^2)^{1/2}\left[1-(A-B z)^2)\right]}{(A-z B)^2}\nonumber\\
&=&\frac{2}{\left(1-z^{2}_0\right)^{3/2}}
\left[q^{(H)}_2(1,A)-q^{(H)}_2\left(z^{}_0,A\right) \right]~.
\eea
Here, $z$ and $z^{}_0$ are the same as in Eq.~(\ref{intI1}), and
\be\l{I2tVzH}
q^{(H)}_2(z,A)=\frac{(z^2-2Az+8)\sqrt{1-z^2}}{2(2A-z)}
-\frac92 \arcsin(z)
+\frac{8A}{\sqrt{1-4A^2}}\ln\left[\zeta(z,A)\right];
\ee
see Eq.~(\ref{arglog}) for $\zeta$.
In Eq.~(\ref{I2tVH}), we used the expression, Eq.~(\ref{rhobMR}),
 for the  
density
$\overline{\rho}$, 
based on the  approximate
outer-inner Schwarzschild stitching 
boundary condition,
Eq.~(\ref{boundcond1}).

  \setcounter{equation}{0}
\renewcommand{\theequation}{D\arabic{equation}}

\section{To calculations of the surface mass,
      energy and
  MI components}
\label{appD}

Let us consider the integral,

\vspace{-0.5cm}
\be\l{Uint}
\mathcal{I}=\int_0^\infty \d r \mathcal{J}(r) Q(r) 
\left(
\frac{\partial \rho}{\partial r}\right)^2~,
\ee
where $\mathcal{J}(r)$ is the volume-shortness factor,
 Eqs.~(\ref{Jin}) and (\ref{Jout}), and
 $Q(r)$ is another assumed
 smooth function of the radial coordinate $r$
as compared to the derivative $\partial \rho/\partial r$.
The 
latter is 
a bell-like function near the ES
 with a sharp maximum at $r=R$ (Appendix \ref{appA1}). 
 The 
function $Q(r)$ is different 
 for the surface components of the energy
$E_S$, mass $M_S$, and MI $\tilde{\Theta}_S$ and
 $\mathcal{T}_S$ components
in Eqs.~(\ref{enertot}), (\ref{MNStot}), and (\ref{adMIVS}),
respectively. 
Taking the assumed smooth functions $\mathcal{J}(r)$ 
and $Q(r)$ off the integral
at $r=R$, and
transforming the integration variable $r$ to $\rho$ by
$\rho=\rho(r)$, $d r=d \rho/(d \rho/dr)$, one can use Eq.~(\ref{eq0})
for the derivative $\partial \rho/\partial r$. As the approximate result
 up to a constant, one has
\be\l{Uint1}
\mathcal{I} \propto \mathcal{J}(R) q(R)
\frac{a}{R}\int_0^{\overline{\rho}} \d \rho 
\sqrt{ 
\epsilon^{}_G}~, 
\ee
where $\overline{\rho}$ is the inner saturation-density limit.
This integral $\mathcal{I}$ 
can be expressed 
in terms of the
tension coefficient
$\sigma \propto a/R$, Eq.~(\ref{sigma1}). Finally, we arrive at the expressions
for the  surface energy $E_S$, Eq.~(\ref{enertots}), 
    the mass $M_S$, Eq.~(\ref{MNSSgen}), and 
    the MI components $\tilde{\Theta}_S$,
   Eqs.~(\ref{adMItS}),  
and
$\mathcal{T}_S$, 
Eq.~(\ref{adMItfS}).

\end{appendices}

\bibliographystyle{iopart-num}

\bibliography{refs}

\end{document}